\documentclass[galaxies,article,accept,moreauthors,10pt,a4paper]{mdpi}

\firstpage{1}
\articlenumber{49}
\doinum{10.3390/galaxies6020049}
\pubvolume{6}
\pubyear{2018}
\copyrightyear{2018}
\history{Received: 31 October 2017; Accepted: 14 April 2018; Published: 18 April 2018}
\updates{yes}

\usepackage{aas_macros}
\usepackage{epsfig, graphicx, subcaption}
\usepackage{longtable}
\usepackage{amsmath, amsfonts, amssymb, mathrsfs}
\usepackage{color}
\usepackage{booktabs}
\usepackage{multirow}
\usepackage{soul}
\usepackage{microtype}
\usepackage{upgreek}

\usepackage{array} \newcommand{\PreserveBackslash}[1]{\let\temp=\\#1\let\\=\temp}
\newcolumntype{C}[1]{>{\PreserveBackslash\centering}m{#1}}
\newcolumntype{R}[1]{>{\PreserveBackslash\raggedleft}m{#1}}
\newcolumntype{L}[1]{>{\PreserveBackslash\raggedright}m{#1}}

\newcommand\figcaption{\def\@captype{figure}\caption}
\newcommand\tabcaption{\def\@captype{table}\caption}

\pdfoutput=1

\Title{Effelsberg Monitoring of a Sample of RadioAstron Blazars: Analysis of Intra-Day Variability}

\Author{Jun Liu $^{1,2,3,}$*, Hayley Bignall $^{4}$, Thomas P. Krichbaum $^{1}$, Xiang Liu $^{2,3}$, Alex Kraus $^{1}$, Yuri~Y.~Kovalev $^{5,6,1}$, Kirill V. Sokolovsky $^{5,7,8}$, Emmanouil Angelakis $^{1}$ and J. Anton Zensus $^{1}$}

\AuthorNames{Firstname Lastname, Firstname Lastname and Firstname Lastname}

\address{%
$^{1}$ \quad Max-Planck-Institut f\"ur Radioastronomie, Auf dem H\"ugel 69, D-53121 Bonn, Germany; tkrichbaum@mpifr-bonn.mpg.de (T.P.K.); akraus@mpifr-bonn.mpg.de (A.K.); yyk@asc.rssi.ru (Y.Y.K.); eangelakis@mpifr-bonn.mpg.de (E.A.); azensus@mpifr-bonn.mpg.de (J.A.Z.)\\
$^{2}$ \quad Xinjiang Astronomical Observatory, CAS, 150 Science 1-Street, Urumqi 830011, China; liux@xao.ac.cn\\
$^{3}$ \quad Key Laboratory of Radio Astronomy, Chinese Academy of Sciences, Urumqi 830011, China\\
$^{4}$ \quad CSIRO Astronomy and Space Science, P.O. Box 1130, Bentley, WA 6102, Australia; hayley.bignall@csiro.au \\
$^{5}$ \quad Astro Space Center of Lebedev Physical Institute, Profsoyuznaya 84/32, 117997 Moscow, Russia; kirx@kirx.net \\
$^{6}$ \quad Moscow Institute of Physics and Technology, Dolgoprudny, Institutsky per., 9,  141700 Moscow, Russia \\
$^{7}$ \quad IAASARS, National Observatory of Athens, Vas. Pavlou \& I. Metaxa, 15236 Penteli, Greece \\
$^{8}$ \quad Sternberg Astronomical Institute, Moscow State University, Universitetskii pr. 13, 119992 Moscow, Russia
}
\corres{Correspondence: jliu@mpifr-bonn.mpg.de}

\abstract{\textls[-15]{We present the first results of an ongoing intra-day variability (IDV) flux density monitoring program of 107 blazars, which were selected from a sample of RadioAstron space very long baseline interferometry (VLBI) targets. The~IDV observations were performed with the Effelsberg 100-m radio telescope at 4.8\,GHz, focusing on the statistical properties of IDV in a relatively large sample of compact active galactic nuclei (AGN). We investigated the dependence of rapid ($<$3 day) variability on various source properties through a likelihood approach. We found that the IDV amplitude depends on flux density and that fainter sources vary by about a factor of 3 more than their brighter counterparts. We also found a significant difference in the variability amplitude between inverted- and flat-spectrum radio sources, with the former exhibiting stronger variations. $\gamma$-ray loud sources were found to vary by up to a factor 4 more than $\gamma$-ray quiet ones, with 4$\sigma$ significance. However a galactic latitude dependence was barely observed, {which suggests that it is predominantly the intrinsic properties (e.g., angular size, core-dominance) of the blazars that determine how they scintillate, rather than the directional dependence in the interstellar medium (ISM).} We showed that the uncertainty in the VLBI brightness temperatures obtained from the space VLBI data of the RadioAstron satellite can be as high as $\sim$70\% due to the presence of the rapid flux density variations. Our statistical results support the view that IDV at centimeter wavelengths is predominantly caused by interstellar scintillation (ISS) of the emission from the most compact, core-dominant region in an AGN.}}


\begin{document}


\section{Introduction}

RadioAstron is an international collaborative mission with a 10-m radio telescope onboard the SPEKTR-R spacecraft launched in July 2011 \cite[]{Kovalev2012,Sokolovsky2013, Kovalev2014}. The~space telescope observes at wavelengths of 92\,cm (324\,MHz, P-band), 18\,cm (1.7\,GHz, L-band), 6\,cm (4.8\,GHz, C-band), and 1.3\,cm (22.2\,GHz, K-band), forming space very long baseline interferometry (space VLBI, or SVLBI) together with ground-based radio telescopes, with the highest angular resolution achievable (e.g., $\sim$7 $\mu$as at the K-band) so far.

One of the Key Science Projects (KSPs) of the space mission is to investigate
the extremely high brightness temperature of active galactic nuclei (AGNs) with the unprecedented long baselines of up to 28 Earth diameters, which will significantly improve our understanding on the mechanisms of AGN radio emissions close to the central engine. Coordinated ground-based flux density monitoring of the RadioAstron targets at centimeter wavelengths is essential to estimate the effect of interstellar scintillation (ISS) on the SVLBI visibilities. As for an ISS `screen' within a few hundred parsecs from the Sun, the characteristic scale of scintillation pattern is comparable to the length of the RadioAstron VLBI baseline.

In order to measure the magnitudes and timescales of ISS of blazars observed by RadioAstron, in 2014 we started a RadioAstron target-triggered flux monitoring program with the Effelsberg 100-m radio telescope at 4.8\,GHz. The~single dish ISS monitoring and space VLBI with RadioAstron offer independent probes of the structure of blazar `cores' at microarcsecond angular scales. Direct measurements of the sizes of scintillating sources with RadioAstron help to determine properties of the interstellar scattering screens, such as their distance and scattering strength. In turn, the focusing and defocusing effects of ISS may have a significant influence on the measured space VLBI visibilities, and it is essential to understand these effects for a complete analysis of the RadioAstron data. Thus,~these two probes of microarcsecond-scale structure are highly complementary.

\textls[-15]{In the cm regime, rapid flux density variations at timescales of a day or less, known as intra-day variability (IDV, \cite[][]{Witzel1986, Heeschen1987}), are frequently observed in compact flat-spectrum radio sources. IDV is present in a significant fraction ($\sim$20--50\%) of flat-spectrum radio sources (e.g., quasars and BL Lac objects) \cite{Quirrenbach1992, Lovell2008}. The~physical mechanism responsible for such variability remains open for debate, with models involving both source-extrinsic and -intrinsic explanations. In many cases the IDV phenomenon is explained by scattering of radio waves by turbulent ionized structures in the Milky Way (e.g.,~\citep[][]{Kedziora-Chudczer1997,Jauncey2001, Dennett-Thorpe2002, Bignall2003, Bignall2006, Liu2013}). On the other hand, some observational evidence, such as large polarization swing, frequency dependence of IDV amplitude, multi-frequency correlation/anti-correlation, and intermittent IDV with structural change, etc., demands a source-intrinsic origin (e.g., \citep[][]{Qian2004, Krichbaum2002, Wagner1996, Liu2015a}).}

However, if interpreted as being source-intrinsic, the size of IDV emitting region---through causality and light-travel time arguments---should be less than tens of $\mu$as. This will lead to very high apparent brightness temperature ($T_B$) that is near or, in many cases, several orders of magnitude in excess of $10^{12}$ K the inverse-Compton (IC) limit \citep{Kellermann1969}. Thanks to the unprecedented long baseline of the RadioAstron space VLBI, we are able to study the $T_B$ of blazar cores up to 10$^{15}$--10$^{16}$ K. In fact, the~RadioAstron AGN survey program has already discovered $T_B$ well in excess of the inverse-Compton limit in some sources, e.g.,~$T_B\sim10^{14}$ K for 3C273 \cite{Kovalev2016}. The~cause of the excess---due to high Doppler boosting or a violation of the inverse-Compton limit---remains essentially undetermined. Recent study suggests it may arise from refractive substructure introduced by scattering in the ISM \cite{Johnson2016}.

In this paper, we present a statistical analysis of the IDV of our sample as the first results of the program. We test the dependence of IDV on various source properties and discuss the implications on the origin of IDV.

\section{Sample Selection, Observations and Data Reduction}

So far, five observing sessions have been carried out at 4.8\,GHz. For each session, the main targets were chosen from the RadioAstron block schedule\footnote{\url{http://www.asc.rssi.ru/radioastron/schedule/sched.html}}. In order to enable high-precision flux density measurements, a nearby non-variable calibrator was selected for each target based on the result of an IDV survey with the Urumqi 25-m radio telescope (\cite[][]{Liu2018}, in preparation) as well as a variability survey conducted with the Very Large Array (MASIV survey,~\cite{Lovell2003}). Note that both these two surveys were performed at frequencies close to 4.8\,GHz. A few sources of particular interest were occasionally added to the list as well. With this procedure of source selection, the final source number in each session is $\sim$40, and the total number of sources observed in the whole campaign is 112. However for the statistical analysis in present work, we remove five~steep-spectrum calibrators (0836 + 710, 0951 + 699, 3C286, 3C48 and NGC7027) which were observed in all epochs only for calibration purposes. This leaves us with a sample of 107 sources.

Since all the sources are point-like to the beam of the Effelsberg radio telescope at 4.8\,GHz, the observations were performed in cross-scan mode, where the antenna beam pattern was driven orthogonally over the source position, stacking four sub-scans for reaching the desired sensitivity. A duty cycle consisted of the observation of the target sources as well as their nearby non-variable secondary calibrators. The~average duty cycle is 0.36 h$^{-1}$ in the campaign, which translates into an average time sampling of $\sim$2.8 h for each source.

The~basic observing information is summarized in Table~\ref{tab:obs_info}. In column 1 to 7 we report the epoch designation, starting and ending date, duration, number of sources observed, mean number of flux density measurements per hour, average number of measurements per hour for each source (duty cycle, which represents the shortest time scale on which we can search for IDV), and the average raw modulation index of calibrators which characterizes the systematic uncertainty (see definition in Section~\ref{sec:def_m}), respectively.

\begin{table}[H]
\centering
\caption{Basic information for the five epochs of observing sessions.}
\label{tab:obs_info}
\scalebox{0.85}[0.85]{
  \begin{tabular}{C{1.5cm}C{2cm}C{2cm}C{3cm}C{3cm}C{2cm}C{1.5cm} }
  \toprule
\textbf{Epoch} & \textbf{Date}  & \textbf{Duration [\boldmath {$h$}]}   & \textbf{Number of Observed Sources}  & \textbf{Average Sampling [\boldmath {$h^{-1}$]}}  & \textbf{Duty Cycle [\boldmath {$h^{-1}$}]}    & \boldmath {$m_c$ \textbf{ [\boldmath {\%}]}}  \\
\midrule
A & 18.07--20.07.2014    & 62.0   & 37    & 14.8    & 0.40    & 0.50  \\
B & 12.09--15.09.2014     & 66.6   & 45    & 15.9    & 0.35    & 0.40  \\
C & 31.07--06.08.2015     & 73.6   & 42    & 14.3    & 0.34    & 0.40  \\
D & 17.12--21.12.2015     & 82.4   & 39    & 14.5    & 0.37    & 0.35  \\
E & 20.12--24.12.2016     & 84.4   & 41    & 14.0    & 0.34    & 0.60  \\
\bottomrule
\end{tabular}}
\end{table}

Frequent switching between targets and calibrators allows us to monitor the antenna gain variations with elevation and time, thus improving the subsequent flux density calibration. The~data calibration was done in the well-established standard manner, and enabled us to achieve a high precision of flux density measurements (see, e.g.,~\cite[][]{Kraus2003}). As the first step of the data calibration, a Gaussian profile is fitted to each sub-scan. The~amplitude of the Gaussian is a measure of the source strength, expressed in units of antenna temperature. After applying a correction for small pointing offsets, the amplitudes of all individual sub-scans in one cross-scan are averaged. Subsequently we correct the measurements for the elevation-dependent gain of the antenna and systematic time-dependent effects, using the secondary calibrators close to target sources. Finally, the measured antenna temperature is converted to absolute flux density by a scaling factor determined by utilizing the frequently observed primary calibrators 3C286, 3C48, and NGC7027 \citep{Baars1977, Ott1994}.

The~overall error on a single measurement is derived from the formal statistical uncertainty and error propagation in each step of data calibration. The~resulting uncertainties usually lie in the range of 0.3--0.7\% of total flux density. The~result of this calibration procedure can be evaluated by measuring the residual scatter, $m_c$, of the calibrators (see definition in Section \ref{sec:def_m}). For most of the observing sessions, the residual scatter is 0.3--0.5\% of total flux density.

\begin{figure}[H]
\centering
     \includegraphics[width=0.8\textwidth]{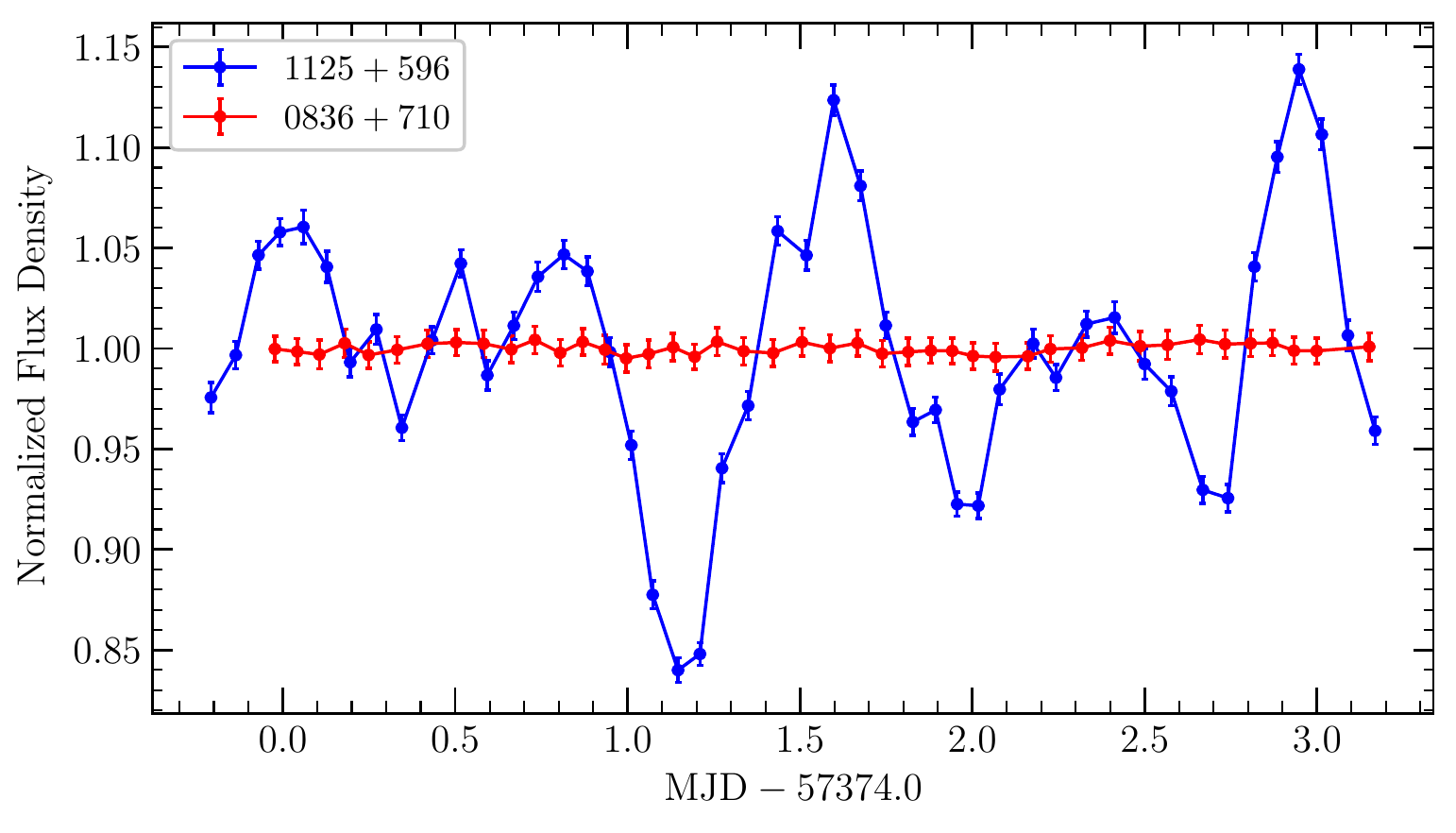}
     \caption{The~intra-day variability (IDV) lightcurve (normalized flux density $S/\langle S \rangle$ versus time) of 1125 + 596 (blue line), and~the calibrator 0836 + 710 (red line) in epoch D.}
     \label{fig:lc_example}
\end{figure}

\textls[-15]{With the data calibration procedure described above, the lightcurves of all the observed sources were obtained. In Figure~\ref{fig:lc_example} we present an example of lightcurve for the target source 1125 + 596 observed in epoch D. The~result of calibrator source 0836 + 710 in the same epoch is superimposed, to demonstrate the stability of the observing system and accuracy of data calibration.}

\section{Variability Parameters}
\label{sec:var_pars}

\textls[-15]{A number of parameters are defined to characterize the IDV. For each light curve the `raw' modulation index $m$, `intrinsic' modulation index $\overline{m}$, $\chi^2$, and reduced $\chi^2$ are derived. Here we give a brief definition and description of these quantities, the reader is referred to ~\cite[][]{Fuhrmann2008, Richards2011} for more details.}

\subsection{Raw Modulation Index} \label{sec:def_m}

 The~raw modulation index is related to the standard deviation of flux density $\sigma$ and the mean value of flux density $\langle S \rangle$ in the time series by
\begin{equation}
    \label{eq:m}
    m[\%]=100\cdot\frac{\sigma}{\langle S \rangle}
\end{equation}
and yields a measure for the strength of observed variations. The~average value of raw modulation index for all the observed calibrators ($m_c$) usually represents the calibration accuracy and characterizes the systematic uncertainty.

\subsection{Intrinsic Modulation Index} \label{sec:def_im}

The~intrinsic modulation index \cite[][]{Richards2011} is an alternative estimator to quantify the variability that would be observed in the absence of measurement errors with ideal time sampling. Note that this use of `intrinsic' is not referring to source-intrinsic variability, but includes any intrinsic and extrinsic (ISS-induced) variations in the received flux density.

With the assumption that the `true' flux densities for each source are normally distributed with mean $S_0$, standard deviation $\sigma_0$, and intrinsic modulation index $\overline{m}=\sigma_0/S_0$, the probability density for the true flux density $S_t$ is
\begin{equation}
    p(S_t, S_0, \sigma_0)=\frac{1}{\sigma_0\sqrt{2\pi}} \exp \left[-\frac{(S_t-S_0)^2}{2\sigma_0^2}\right] \,.
\end{equation}

Furthermore, we assume that the observation process for the $j$th data point adds normally distributed error with mean $S_t$ and standard deviation $\sigma_j$. Then the likelihood for a single observation is given by
\vspace{30pt}
\begin{equation}
 \ell_j(S_0, \sigma_0) = \int_{\,\rm all \,\, S_t} \!\!\!\!\!\!\!\!dS_t \frac{\exp \left[
-\frac{(S_t-S_j)^2}{2\sigma_j^2}\right] }{\sigma_j \sqrt{2\pi}} \frac{\exp \left[
-\frac{(S_t-S_0)^2}{2\sigma_0^2}\right]}{\sigma_o \sqrt{2\pi}} \,,
\end{equation}
which after combining $j=1,...N$ measurements and substituting $\overline{m}S_0=\sigma_0$, gives
\begin{equation} \label{eq:likelihood}
    \mathcal{L}(S_0, \overline{m}) = S_0\left(\prod_{j=1}^N{\frac {1}{\sqrt{ 2\pi \left(
                \overline{m}^2{S}_0^2+\sigma_j^2 \right)}}}  \right) \exp \left(-\frac {1}{2} \sum
_{j=1}^N{\frac {\left( S_j-S_0 \right)^2}{\overline{m}^2{S}_0^2+\sigma_j^2}}\right) \,.
\end{equation}

By maximizing the joint likelihood given by Equation~(\ref{eq:likelihood}), we find our estimates of $S_0$ and $\overline{m}$.

\begin{figure}[H]
\centering
     \includegraphics[width=0.45\textwidth]{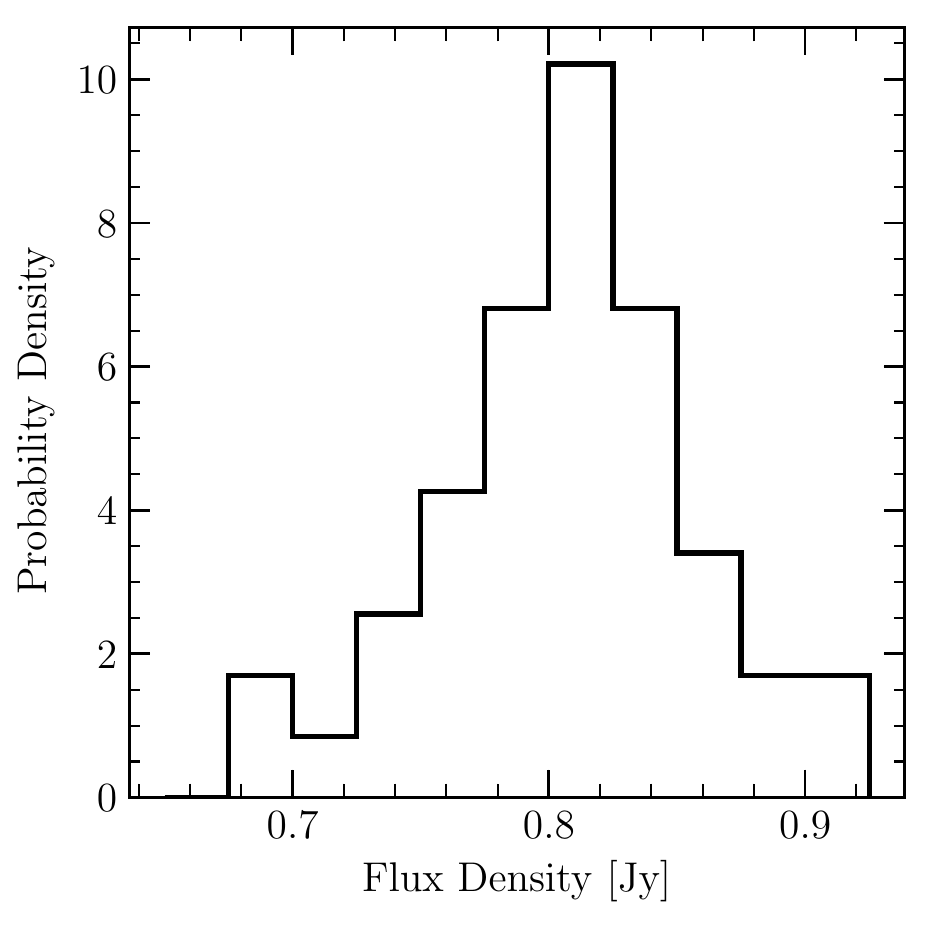}
     \includegraphics[width=0.45\textwidth]{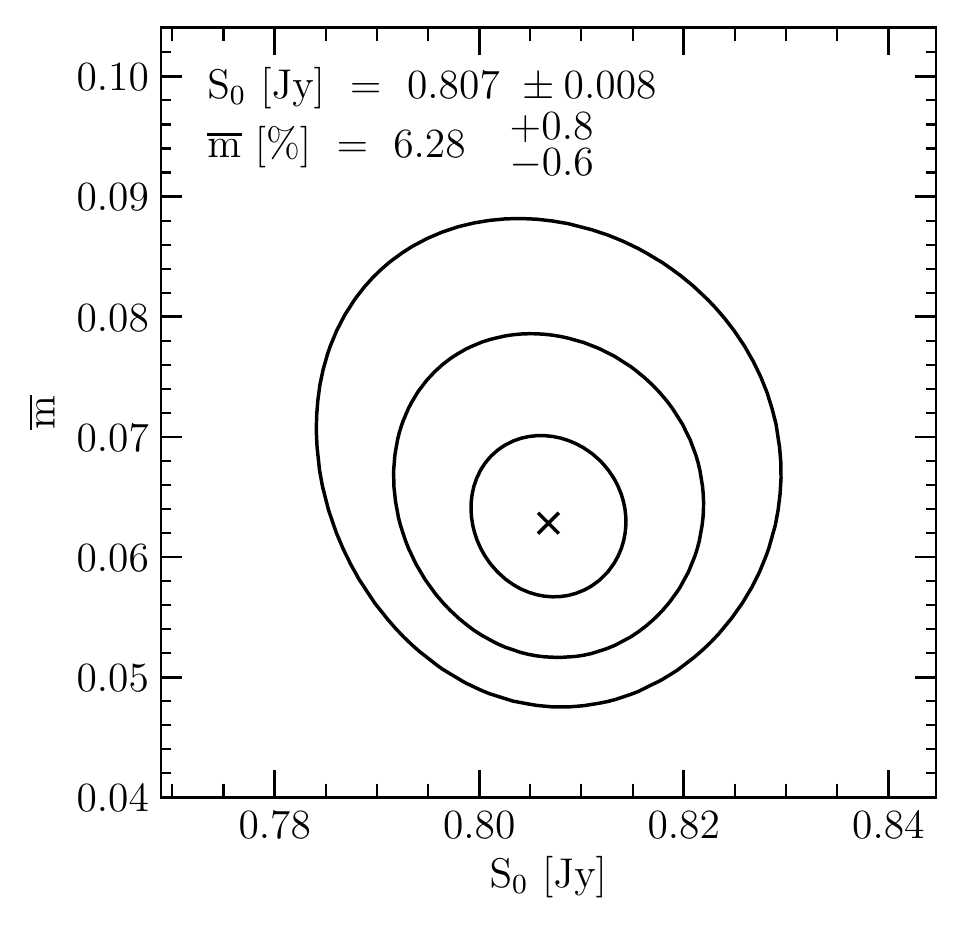}
     \caption{Maximum likelihood estimation for blazar 1125 + 596 observed in epoch D. (\textbf{Left}) Distribution of measured flux density. (\textbf{Right}) 1$\sigma$, 2$\sigma$, and 3$\sigma$ contours of the joint likelihood $\mathcal{L}(\overline{m}, S_0)$.}
     \label{fig:mle_example}
\end{figure}

The~maximum-likelihood we applied makes the assumption that distribution of flux densities from a source is distributed normally. For many sources, this is a good description of the data. As an example, in the left panel of Figure~\ref{fig:mle_example}, we plot the histogram of epoch D data set flux densities from 1125 + 596 (for which the IDV curve is shown in Figure~\ref{fig:lc_example}). It is clear that the histogram approximately forms a Gaussian profile. In the right panel of Figure~\ref{fig:mle_example}, we plot the most likely values and the $1\sigma$, $2\sigma$, and $3\sigma$ isolikelihood for the same source. The~contours were computed to contain 68.26\%, 95.45\%, and 99.73\% of the volume beneath the likelihood surface. In this way, we obtained the most likely values of $\overline{m}$ and $S_0$, as well as their $1\sigma$ uncertainties. We note that a rigorous estimate of the uncertainty in individual $\overline{m}$ is essential for evaluating the significance of differences between $\overline{m}$ through population studies which will be demonstrated in Section~\ref{sec:pop_comp}.

\subsection{\texorpdfstring{$\chi^2$}{chi2} and Reduced \texorpdfstring{$\chi^2$}{chi2_r}} \label{sec:def_chi}

Finally, as a criterion to identify the presence of variability, the null-hypothesis of a constant function is examined via a $\chi^2$-test
\begin{equation}
\chi^2 = \sum _{ j=1 }^{ N }{ { \left( \frac {S_j - \left< S \right>  }{ \sigma_j }
\right)  }^{ 2 } }
\end{equation}
and the reduced value of $\chi^2$
\begin{equation} \chi_r^2 = \frac { 1 }{ N-1 }\sum _{ j=1 }^{ N }{ { \left( \frac {S_j - \left< S \right> }{ \sigma_j } \right)  }^{ 2 } } \end{equation}

A source is classified as variable if the $\chi^2$-test gives a probability of <0.01$\%$ for the assumption of constant flux density (99.99\% significance level for variability).

\section{Statistical Results} \label{sec:stat_overall}

Table~\ref{tab:all_result} summarizes the basic properties and statistical results of sources in the sample. Columns 1 and 2 give the source name and epoch designation. Observation results are presented in columns 3, 4, and 5. Column 6 gives variability classification. A `+' is given if the source is identified as variable. Source-basic properties, i.e., galactic latitude, spectral index, redshift, VLBI core size at 5\,GHz, and core-dominance, are shown in column 7--11, respectively. Column 12 lists the $\gamma$-ray loudness. A '+' is marked if the source is included in the Fermi Large Area Telescope Third Source Catalog \cite[3FGL,][]{Acero2015}. The galactic latitudes and redshifts were taken from the NASA/IPAC Extragalactic Database (NED\footnote{\url{https://ned.ipac.caltech.edu/}}) and spectral indexes (defined by $S\propto\nu^{\alpha}$ where $S$ the flux density and $\nu$ the observing frequency) from the Radio Fundamental Catalog\footnote{\url{http://astrogeo.org/rfc/}}. The~VLBI core sizes were extracted from~\citep{Pushkarev2015}. The~core-dominance, defined by $f_{\rm c}=S_{\rm core}/S_{\rm total}$, was derived by use of data from the Radio Fundamental Catalog.

In the following we overview the statistical properties of the sample. We present the distributions of source flux density, spectral index, galactic latitude, and intrinsic modulation index. For the purpose of population study, the distribution of $\overline{m}$ is analytically modeled.

\subsection{Variability Classification}

\begin{figure}[H]
\centering
    \includegraphics[width=0.55\textwidth]{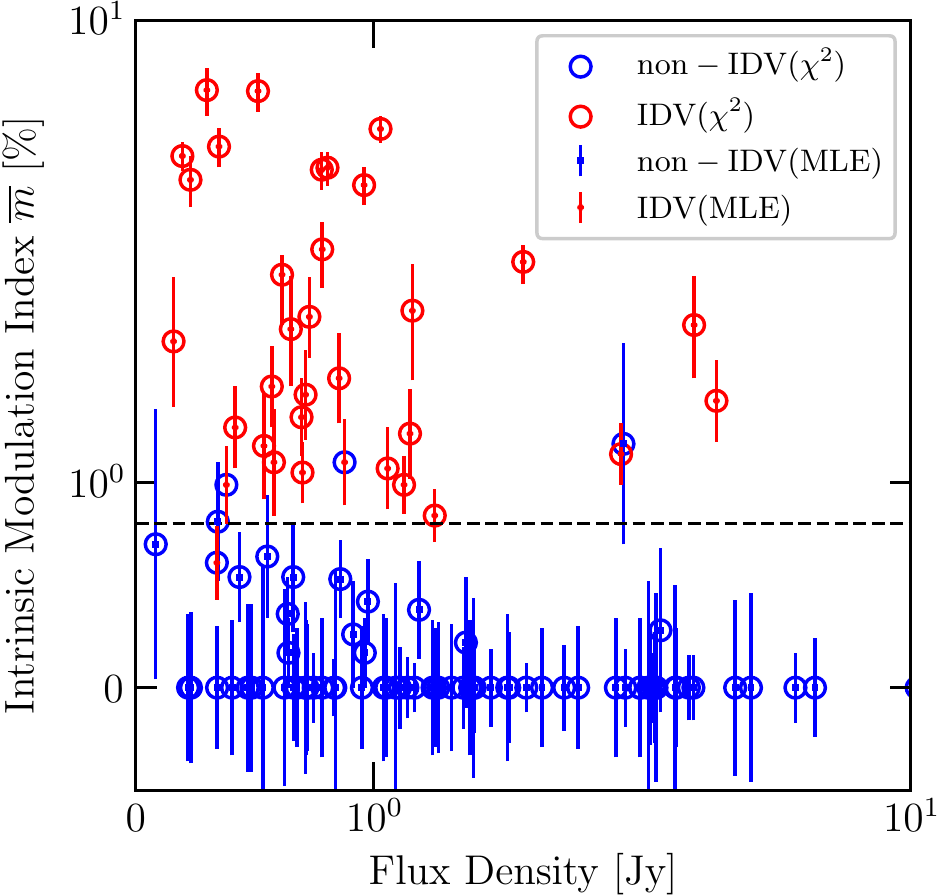}
    \caption{Intrinsic modulation index $\overline{m}$ plotted against mean flux density. Sources classified as variables are plotted in red while non-variables are in blue. Classification with $\chi^2$-test ($\chi^2$) plotted as circles while maximum likelihood estimation (MLE) is shown as an error bar.}
    \label{fig:idv_class}
\end{figure}

To verify the presence of IDV, a $\chi^2$-test is applied to each observed lightcurve. Of the 107 targets observed, 31 sources are found to exhibited IDV in at least one observing epoch, while the rest of the sample does not reveal evident IDV in any epoch. This leads to an IDV detection rate of $\sim$30\% for current sample, comparable to earlier studies~\cite{Quirrenbach1992, Lovell2008}.
In order to minimize the chance of misclassification, a cross-check of the intrinsic modulation index is performed alternatively. If the maximum-likelihood $\overline{m}$ is less than $3\sigma$ away from $\overline{m}$ = 0, we consider that significant variability cannot be established in the source. As a result, the new approach classifies three extra sources as variable, besides the 31 IDV sources previously identified by $\chi^2$ tests. The~result demonstrates that our variability verification with different approaches are mostly consistent with each other. In Figure~\ref{fig:idv_class} we show a scatter plot of the intrinsic modulation index $\overline{m}$ of all sources against their mean flux density, using different symbols for variables and non-variables identified with both approaches. A dashed horizontal line at $\overline{m}$ = 0.75\% roughly separates the variable from non-variable classifications, confirming the result of the $\chi^2$-test.

\subsection{Sample Properties}

\begin{figure}[H]
\centering
    \includegraphics[width=0.96\textwidth]{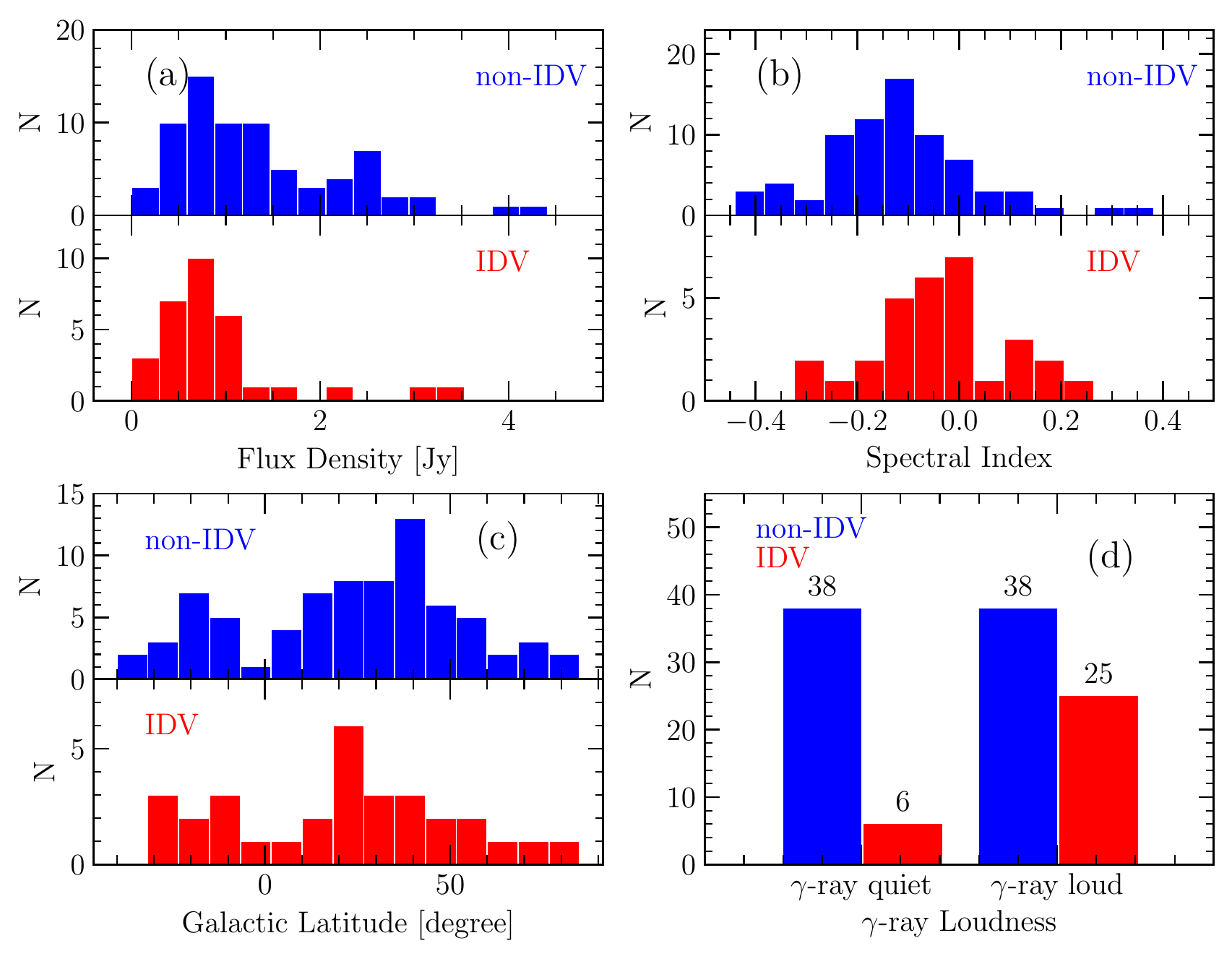}
    \caption{Sample properties. Panels (\textbf{a}--\textbf{d}) present the distribution of flux density, spectral index, galactic latitude, and $\gamma$-ray loudness, respectively. In each panel, IDV sources are plotted in red and non-IDVs in blue.}
    \label{fig:sample_dist}
\end{figure}

\textls[-15]{An overview of the sample properties is presented in Figure~\ref{fig:sample_dist}. In each panel, IDV sources are plotted in red and non-IDVs in blue. The~distribution of flux density is plotted in panel (a), showing a bimodal profile for non-IDVs. It is obvious that IDV sources are mostly clustered at the lower flux density peak, while the higher flux density peak is predominantly occupied by the non-IDV population. Panel (b) shows the unimodal distribution of spectral index. The~variables are well distributed at $\alpha$ > $-$0.1, indicating deviation from the non-variable population. In panel (c) the source galactic latitude shows a distribution peaked at |b|$\sim$35$^{\circ}$. The~occurrence of IDV and non-IDV sources reveals a similar trend, and no clear difference can be visually observed between these two populations. We note that sources with |b|$<10^{\circ}$ are very rare in our sample. The~occurrence of IDV/non-IDV among $\gamma$-ray loud and $\gamma$-ray quiet subsamples is compared in panel (d). Of the 63 sources with a GeV detection by Fermi, 25~exhibit IDV, indicating that $\sim$40\% of the $\gamma$-ray loud sources are variable at cm-wavelength. By contrast the ratio is as low as  $\sim$14\% for $\gamma$-ray quiet sources, indicating a higher occurrence rate of IDV in $\gamma$-ray loud population.}

\unskip

\subsection{Intrinsic Modulation Index \texorpdfstring{$\overline{m}$}{m}}
\label{sec:stat_im}

\begin{figure}[H]
\centering
     \includegraphics[width=0.5\textwidth]{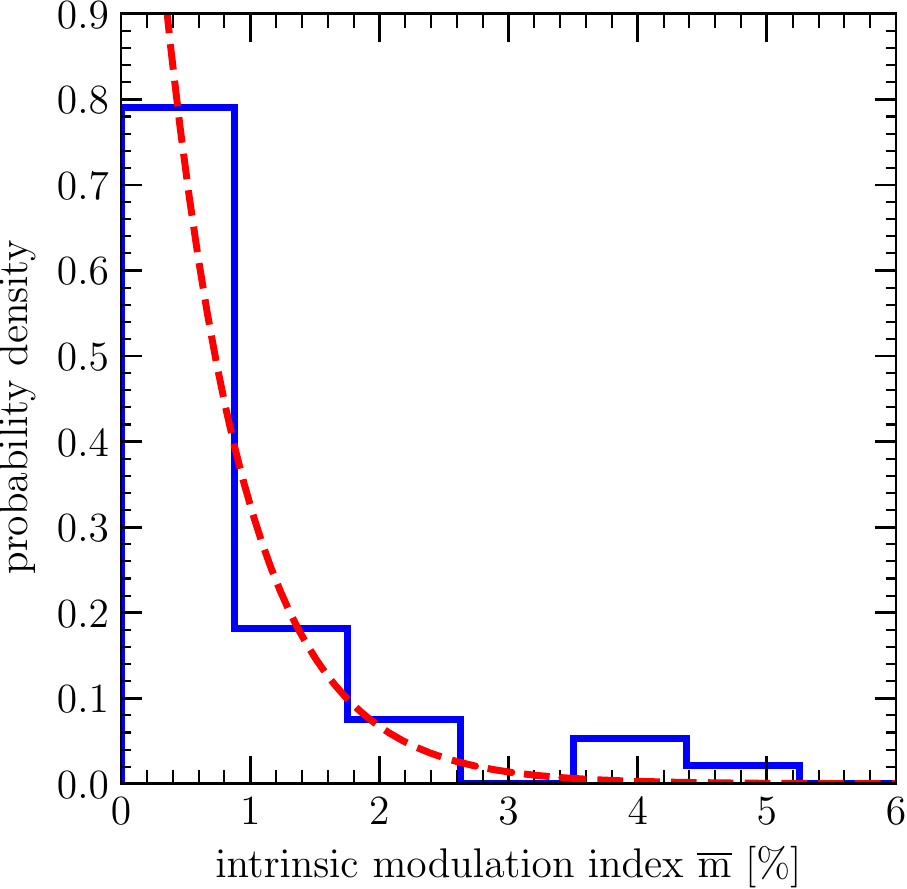}
     \caption{Probability density of the maximum-likelihood intrinsic modulation indexes $\overline{m}$. The~red dashed line represents an exponential distribution with $m_0$ = 0.63\%.}
     \label{fig:dist_im}
\end{figure}

The~probability density of $\overline{m}$ for our monitoring sample is plotted in Figure~\ref{fig:dist_im}. As in \cite{Richards2011, Richards2014}, we~use a monoparametric exponential family of distributions
\begin{equation}
    \label{eq:psd_im}
    f(\overline{m})d\overline{m} = \frac{1}{m_0}\,\exp(-\frac {\overline{m}}{m_0})d\overline{m}
\end{equation}
with mean $m_0$ to model the observed probability density of $\overline{m}$. The~red line represents the best fit with mean $m_0$ = 0.63\% in the form of Equation~(\ref{eq:psd_im}). The~model, which provides an excellent description of the data, will be used to characterize various subgroups of our sample, as we will see in the next section. We note that a median value of $\overline{m}$ is adopted if the source was observed in multiple epochs. The~robustness of the population comparisons presented in Section~\ref{sec:pop_comp} will be discussed in Section~\ref{sec:discuss}.

\section{Population Comparisons} \label{sec:pop_comp}

We now investigate how the variability amplitude, as quantified by $\overline{m}$, depends on source properties, i.e., flux densities, spectral indexes, $\gamma$-ray loudness, and galactic latitudes. To this end, we determine the distribution of $\overline{m}$ for various subsets of our sample with, again, a likelihood maximization method as in \cite{Richards2011}.

\textls[-15]{The~likelihood analysis requires a parent distribution for $\overline{m}$. As shown in Section~\ref{sec:stat_im}, an~exponential distribution {as given in Equation~(\ref{eq:psd_im})}, is a qualitatively reasonable fit to the observed distribution of modulation indexes in our sample. {Under the assumption that the exponential distribution (Equation~(\ref{eq:psd_im})) is the correct underlying distribution for $\overline{m}$ in any given subset (i.e., population) of our source sample, we~can use the parameter $m_0$ as a measure for the population to show strong IDV. To find $m_0$, we perform a likelihood analysis, as described in Section~\ref{sec:pop_method}. We~hereby use the maximum of the normalized likelihood function to determine $m_0$ for each population, or rather we investigate the probability distribution of any possible $m_0$ values to estimate the statistical significance of differences in $m_0$ found for different~populations.}}

We also performed {$k$-sample} Anderson--Darling (A-D) tests \cite{Scholz1987} to each pair of subsamples for crosschecking. The~A-D test is a nonparametric statistical procedure testing the hypothesis that the populations from which two or more samples of data were drawn to be identical. It is a modification of the Kolmogorov--Smirnov (K-S) test and gives more weight to the tails than does the latter, thus~allowing a more sensitive test for {exponential-like} distribution and offering a better option for the statistics in current study.

\subsection{Likelihood Analysis}
\label{sec:pop_method}

Here we briefly introduce the methodology of maximum-likelihood used for population studies in this work. More details can be found in Section 6.3.3 of \cite{Richards2011} where the formalism is well demonstrated.

For a source $i$, the likelihood of observing a modulation index $\overline{m_i}$ with a Gaussian uncertainty $\sigma_i$ drawn from an exponential distribution with mean $m_0$ is
\begin{equation}
    \begin{aligned}
    \ell_i(m_0) & = \int_{\overline{m}=0}^\infty\,d\overline{m} \frac{1}{m_0} \exp\left(-\frac{\overline{m}}{m_0}\right) \frac{1}{\sigma_i\sqrt{2\pi}} \exp \left[-\frac{(\overline{m}-\overline{m_i})^2}{2\sigma_i^2}\right] \\
    & = \frac{1}{m_0\sigma_i\sqrt{2\pi}} \exp \left[ -\frac{\overline{m_i}}{m_0} \left( 1-\frac{\sigma_i^2}{2m_0\overline{m_i}} \right) \right] \times
    \int_{\overline{m}=0}^\infty \exp \left[ -\frac{[\overline{m}-(\overline{m_i}-\sigma_i^2/m_0)]^2}{2\sigma_i^2} \right]\,d\overline{m} \, ,
    \end{aligned}
\end{equation}
where, to obtain the second expression, we have completed the square in the exponent of the integrand. The~last integral can be calculated analytically, yielding
\begin{equation}
    \ell_i(m_0) = \frac{1}{2m_0} \exp \left[ -\frac{\overline{m_i}}{m_0} \left(1-\frac{\sigma_i^2}{2m_0\overline{m_i}}\right) \right] \times
    \left\{1+\mathrm{erf} \left[\frac{\overline{m_i}}{\sigma_i\sqrt{2}} \left(1-\frac{\sigma_i^2}{m_0\overline{m_i}}\right) \right] \right\} \, .
\end{equation}

The~likelihood of $N$ observations of this type is
\begin{equation} \label{eq:pop_likelihood}
    \mathcal{L}(m_0) = \prod_{i=1}^N \ell_i(m_0).
\end{equation}

{
The~probability density function (PDF) of $m_0$ is the normalization of $\mathcal{L}(m_0)$,
\begin{equation} \label{eq:pop_pdf}
	\mathrm{pdf}(m_0)=\frac{\mathcal{L}(m_0)}{\int_{0}^{\infty}\mathcal{L} (m_0)\,d m_0}
\end{equation}

From the maximization of Equation~(\ref{eq:pop_pdf}) we obtain the maximum-likelihood value of $m_0$. The~statistical uncertainty (1$\sigma$ error) on this value can also be obtained by locating the isolikelihood $m_0$-values $m_{01}$ and $m_{02}$ for which
\begin{equation}
	\mathrm{pdf}(m_{01})=\mathrm{pdf}(m_{02})
\end{equation}
and
\begin{equation} \label{eq:conf_interval}
	\frac{\int_{m_{01}}^{m_{02}}\mathrm{pdf}(m_0)dm_0}{\int_{0}^{\infty}\mathrm{pdf}(m_0)dm_0}=0.6826.
\end{equation}

The~confidence intervals are derived in a similar way by substituting the right-hand side of Equation~(\ref{eq:conf_interval}) by, e.g., 0.9545 and 0.9973 for the cases of 2$\sigma$ and 3$\sigma$, respectively. We note that the 1$\sigma$ errors and confidence intervals for the difference of $m_0$ are calculated in the same way.

}\mdseries
With the above introduced formalism, we are able to examine whether the intrinsic modulation index $\overline{m}$ correlates with the properties of the sources in our sample. In the following sections, we will study the distributions of $\overline{m}$-values in the subgroups of our monitoring sample according to some source properties, i.e., flux densities, spectral indexes, $\gamma$-ray loudness, and Galactic latitudes as well.

A summary of population comparisons between various subsamples is tabulated in Table~\ref{tab:pop_result}. For each subsample, in column 1 we list the criteria used for subsample division; in column 2 we present the number of sources in subsample; in column 3 we estimate the most likely values of $m_0$ by maximizing the likelihood function given in Equation~(\ref{eq:pop_pdf}). The associated 1$\sigma$ uncertainties are also provided. For each pair of subsamples compared, in column 4 we calculate the most likely value along with the corresponding 1$\sigma$ uncertainties for the difference in $m_0$; in column 5 we list the significance of the difference. Finally, the $p$-value and significance as estimated from A-D tests are reported in Column 6 and 7, respectively. As shown in the table, the results from two statistical approaches are highly consistent with each other.

\begin{table}[H]
\centering
\caption{Results of population comparisons.}
\label{tab:pop_result}
  \begin{tabular}{ccccrcc}
  \toprule
     & & \multicolumn{3}{c}{\textbf{Likelihood Analysis}} & \multicolumn{2}{c}{\textbf{Anderson--Darling Test}} \\
   \midrule
\textbf{Subsample} & \textbf{Source Num.} & \boldmath{$m_0$}\textbf{ [\boldmath{\%}]}   & \boldmath{$\Delta\,m_0$}  \textbf{[\boldmath{\%}] } & \textbf{Significance}    & \boldmath{$p$} & \textbf{Significance}\\
\midrule
$S_{4.8}<$\,1\,Jy    &53 & $1.24_{-0.165}^{+0.198}$ & \multirow{2}{1.6cm}{$+0.78_{-0.187}^{+0.205}$}  & \multirow{2}{1cm}{$5\sigma$}  & \multirow{2}{1.6cm}{$1.32\times10^{-5}$} & \multirow{2}{0.4cm}{$4\sigma$}    \\
$S_{4.8}\geq$\,1\,Jy &54 & $0.46_{-0.064}^{+0.077}$ &  &   \\
\midrule
$\alpha<$\,$-$0.1       &51 & $0.59_{-0.084}^{+0.102}$ & \multirow{2}{1.6cm}{$-0.47_{-0.189}^{+0.172}$} & \multirow{2}{1cm}{$3\sigma$}  & \multirow{2}{1.6cm}{$5.23\times10^{-4}$} & \multirow{2}{0.4cm}{$3\sigma$}      \\
$\alpha\geq$\,$-$0.1    &56 & $1.07_{-0.139}^{+0.167}$ &  &   \\
\midrule
$\gamma$-ray quiet    &44 & $0.43_{-0.067}^{+0.082}$ & \multirow{2}{1.6cm}{$-0.69_{-0.179}^{+0.162}$} & \multirow{2}{1cm}{$4\sigma$}  & \multirow{2}{1.6cm}{$4.37\times10^{-4}$} & \multirow{2}{0.4cm}{$3\sigma$}      \\
$\gamma$-ray loud     &63 & $1.13_{-0.139}^{+0.164}$ &  &   \\
\midrule
$|b|<\,20^{\circ}$    &34 & $0.98_{-0.163}^{+0.206}$ & \multirow{2}{1.6cm}{$+0.18_{-0.194}^{+0.229}$} & \multirow{2}{1.2cm}{<1$\sigma$}  & \multirow{2}{1.6cm}{$9.67\times10^{-1}$} & \multirow{2}{0.9cm}{<1$\sigma$}      \\
$|b|\geq\,20^{\circ}$ &73 & $0.79_{-0.091}^{+0.107}$ &  &   \\
\bottomrule
\end{tabular}
\end{table}
 \subsection{Flux Density and Redshift} \label{sec:pop_flux}

\textls[-15]{In order to exhibit ISS, a source must contain a component that is sufficiently compact, with~angular diameter comparable to or smaller than the first Fresnel zone of the scattering screen, i.e., on~the order of tens of $\mu$as near a few GHz~(e.g.,~\cite[][]{Bignall2006, Bignall2004, Dennett-Thorpe2000}).}

\begin{figure}[H]
\centering
     \includegraphics[width=0.45\textwidth]{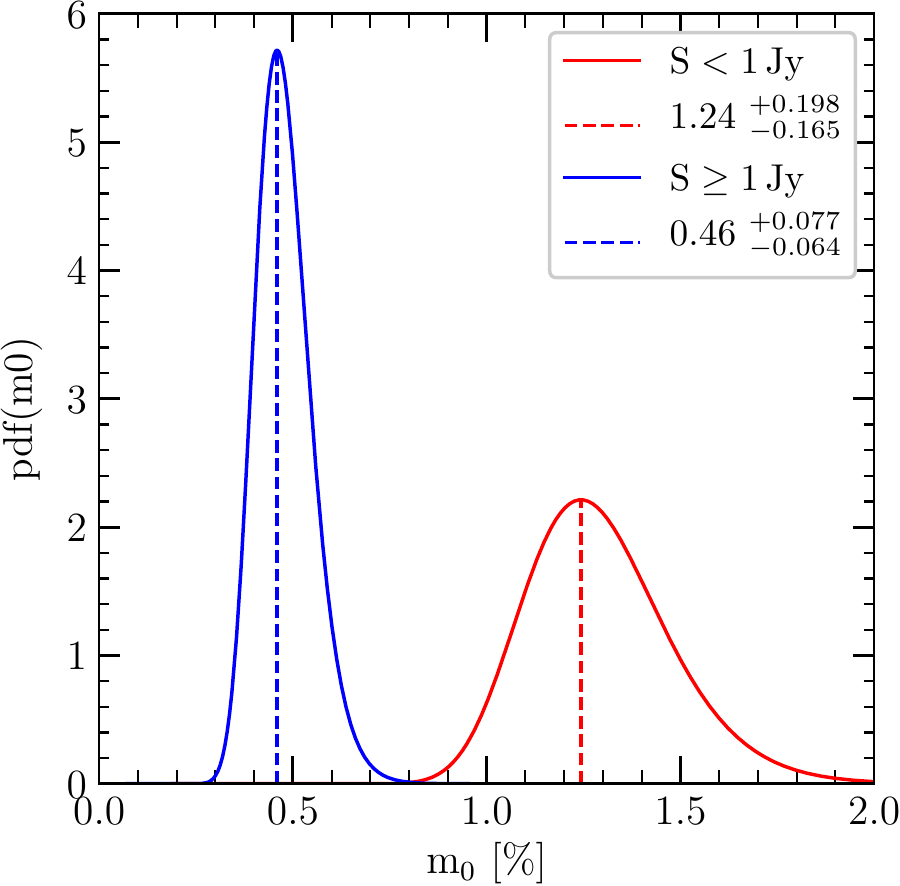}
     \includegraphics[width=0.46\textwidth]{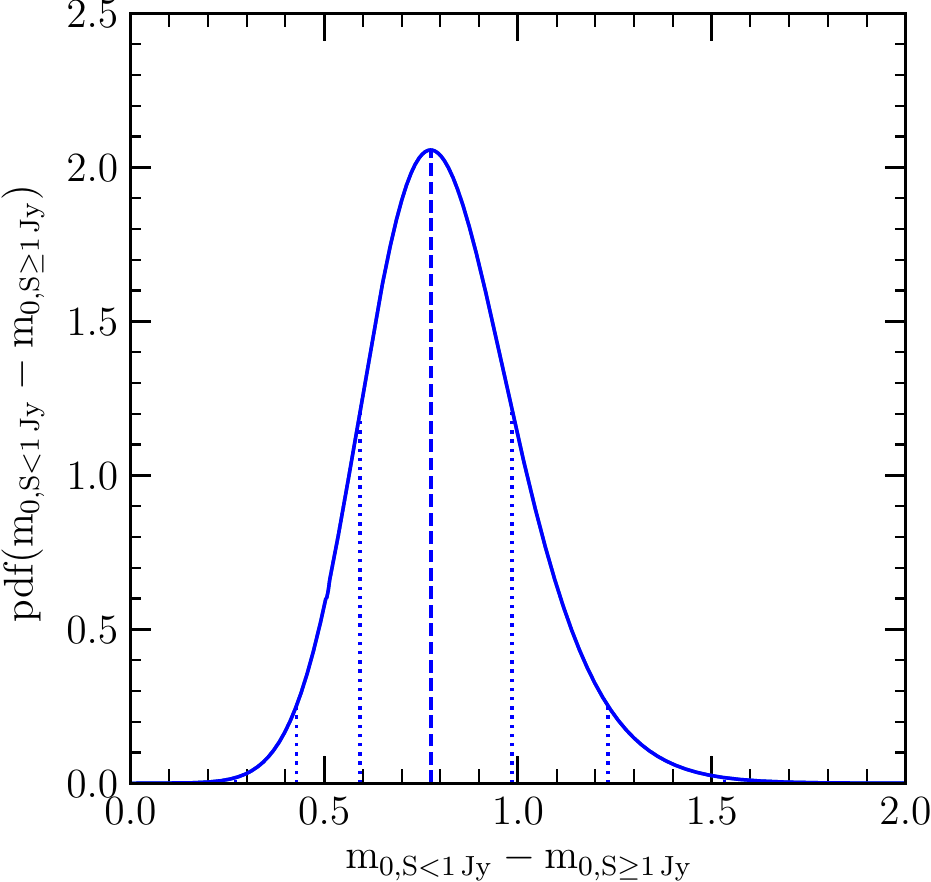}
     \caption{(\textbf{Left panel}) Probability density of $m_0$ for sources with flux density lower (red solid line, maximum-likelihood value and $1\sigma$ error $m_0=1.24^{+0.198}_{-0.165}\%$) and higher (blue solid line, maximum-likelihood value and $1\sigma$ error $m_0=0.46^{+0.077}_{-0.064}\%$) than 1\,Jy in our monitoring sample. The~dashed vertical lines locate the peaks of probability density for the two subsamples. (\textbf{Right panel}) Probability density of the difference between the $m_0$ for the two sets considered in the left panel. The~dashed vertical line shows the peak of the probability density, while the dotted vertical lines represent the 1, 2, and 3$\sigma$ confidence interval. The~peak of the distribution ($0.78^{+0.205}_{-0.187}$) is over $5\sigma$ away from zero.}
     \label{fig:pop_flux}
\end{figure}

We start with testing the dependence of IDV on source flux density, which is tightly bounded with source angular size if the brightness temperature is inverse-Compton limited. A population study is also performed to examine the subsets defined by whether the source flux density is higher or lower than 1\,Jy. The~results of this test are displayed in Figure \ref{fig:pop_flux}. In the left panel, it is obvious that the curves for the two subsamples are not consistent with each other -- weaker sources have, on average, higher IDV amplitude. The~significance of this result is verified by the right panel of Figure \ref{fig:pop_flux}, where we plot the probability density of the difference between the $m_0$ of the two subsets {(which is formally equal to the cross-correlation of their individual distributions).
With the formalism introduced in Section~\ref{sec:pop_method}} the most likely difference is 0.84 percentage of points, which is more than 5$\sigma$ away from zero.
Our~result can be understood in terms of source angular scale. The~angular size of a source can be modeled as a function of its flux density, $S$, and the brightness temperature $T_B$ in source frame, as follows:
\begin{equation}
\label{eq:theta}
\theta=\sqrt{\dfrac{\lambda^2(1+z)S}{2\pi k \delta T_B}},
\end{equation}
where $\lambda$ is the observing wavelength, $z$ is the source redshift, $k$ is the Boltzmann constant, and $\delta$ is the Doppler boosting factor. Therefore, if these sources are limited in brightness temperature, either due to the inverse-Compton catastrophe \cite{Kellermann1969} or energy equipartition (between particles and the magnetic fields \cite{Readhead1994}), the source angular size scales as $\theta\propto S^{0.5}$.
In that case, the brighter sources have larger angular sizes, and thus suppress the ISS. In other words, our finding indicates a source compactness related IDV.
Moreover, with the availability of VLBI data on the core size (column 10 of Table~\ref{tab:all_result}), it is possible to verify the argument by testing the correlation between intrinsic modulation index and VLBI core size at 5\,GHz directly. The~two-tailed, non-parametric Spearman rank correlation test gives a correlation coefficient $r_s=-0.219$ and a $p$-value of $p=6.44\times10^{-2}$. Though it is less significant, a negative trend in the IDV strength over source angular size is implied.

Furthermore, the observed IDV may be also dependent on the source redshift, providing that the sample of sources are both flux density- and brightness temperature-limited \cite{Lovell2008}. In this case the redshift dependence of the angular size is $\theta\propto(1+z)^{0.5}$ due to cosmological expansion. However we do not find a statistically significant relation between redshift and IDV amplitudes in our sample. We note that the scatter in flux density is larger than the scatter in $(1+z)$, and there is no clear relationship seen between source flux density and redshift in this sample. Moreover, a suppression of ISS was observed in the MASIV survey only for sources with $z>2.5$ \cite{Lovell2008}. Our sample has only four sources with $z>2.5$.

\subsection{Spectral Index} \label{sec:pop_alpha}

Early observations showed that scintillating sources tend to have flat or inverted spectra, while~the steep-spectrum radio sources do not scintillate \cite{Heeschen1984}. This can be understood by considering that the flat-spectrum sources are dominated by optically thick, synchrotron self-absorbed components with very high-brightness temperature, and thus most of their flux density is confined to the ultra-compact core region. In contrast, the steep-spectrum sources are dominated by optically thin, less compact components with lower brightness temperatures, often related to an extended VLBI jet.

\begin{figure}[H]
\centering
     \includegraphics[width=0.45\textwidth]{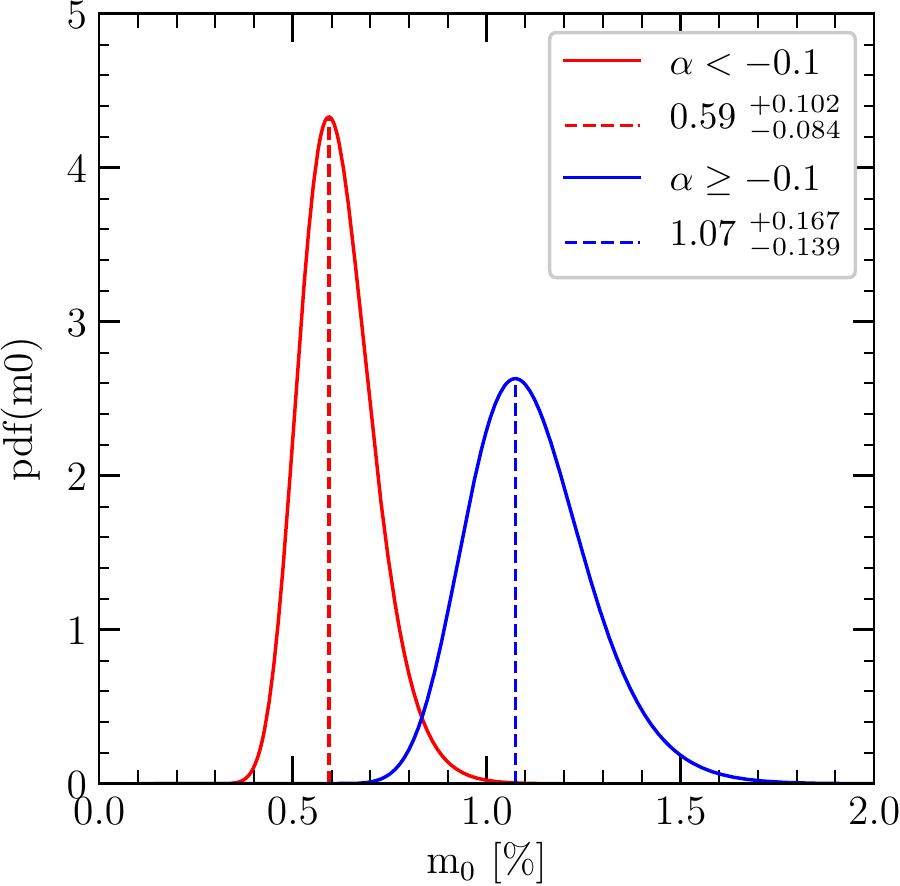}
     \includegraphics[width=0.46\textwidth]{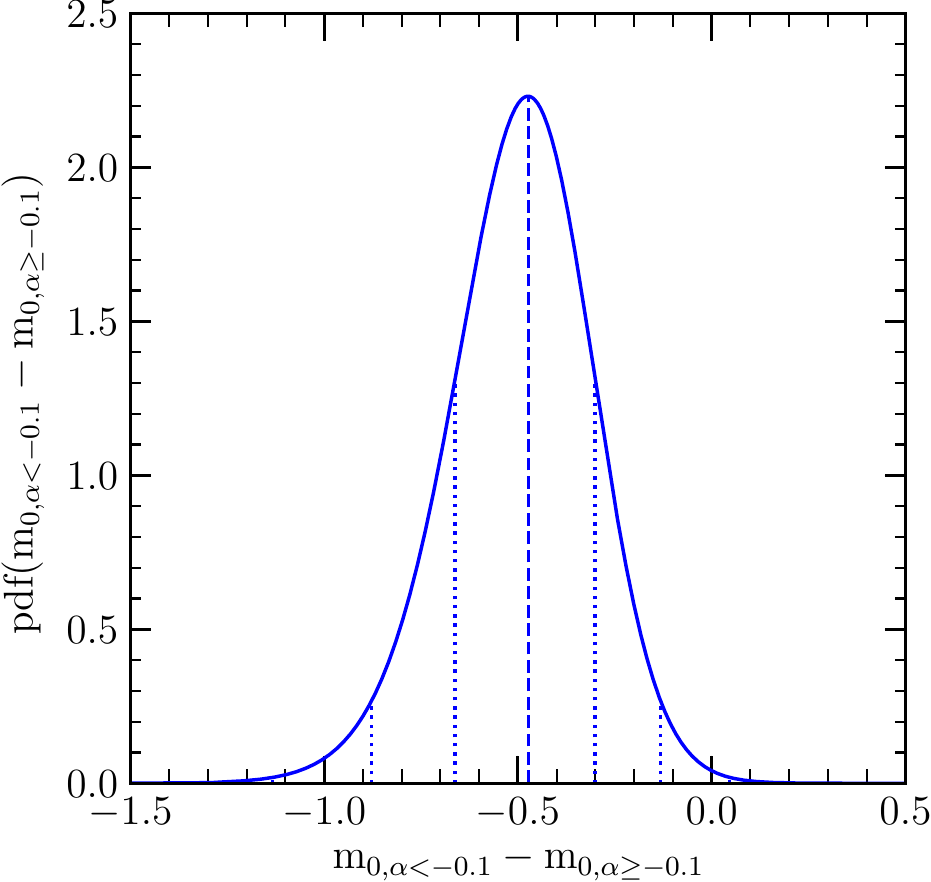}
     \caption{{Similar} to Figure \ref{fig:pop_flux} but for sources with spectral index lower (red solid line) and higher (blue solid line) than $-$0.1. In the left panel the maximum-likelihood value and the associated 1$\sigma$ error are indicated in the legend. In the right panel the peak of the distribution ($-0.47_{-0.189}^{+0.172}$) is over 3$\sigma$ away from zero.}
     \label{fig:pop_alpha}
\end{figure}

To test this argument, we split the sample at $\alpha=-0.1$. This criterion roughly splits our sample into flat and inverted spectra, and produces subsamples of similar numbers of objects. Figure \ref{fig:pop_alpha} depicts the probability distributions of $m_0$ as well as the difference between $m_0$ for the two subsamples. A Spearman rank test between the intrinsic modulation index and source core-dominance (column 11 of Table~\ref{tab:all_result}) confirms the result ($r_s=0.284$ and $p=9.72\times10^{-3}$). The~finding, as anticipated, suggests that sources with inverted spectra are significantly stronger in short-term variability. It has to be noted, however, that the sources in the present sample are mostly compact, core-dominated sources with flat spectrum, unlike the classical steep-spectrum sources reported by \citet{Heeschen1984} which are dominated by their extended emission. Our findings indicate that even within the flat spectrum sources the presence of additional less compact components could in principle reduce their core-dominance, thus~reducing the scintillation.

\subsection{\texorpdfstring{$\gamma$}{gamma}-Ray Loudness} \label{sec:pop_gamma}

The~source of high energy emission is believed to be compact and located close to the central engine of AGNs (e.g., \cite[][]{Kovalev2009, Pushkarev2010}). Relations between the parsec-scale radio properties and $\gamma$-rays of blazars have been intensively investigated since the Fermi $\gamma$-ray Space Telescope was launched (e.g.,~\cite[][]{Lister2009, Ramakrishnan2014, Casadio2015, Hada2016}). In this study we test, through a statistical approach, whether there is a correlation between the $\gamma$-ray loudness and the 4.8\,GHz IDV properties.

\begin{figure}[H]
\centering
     \includegraphics[width=0.45\textwidth]{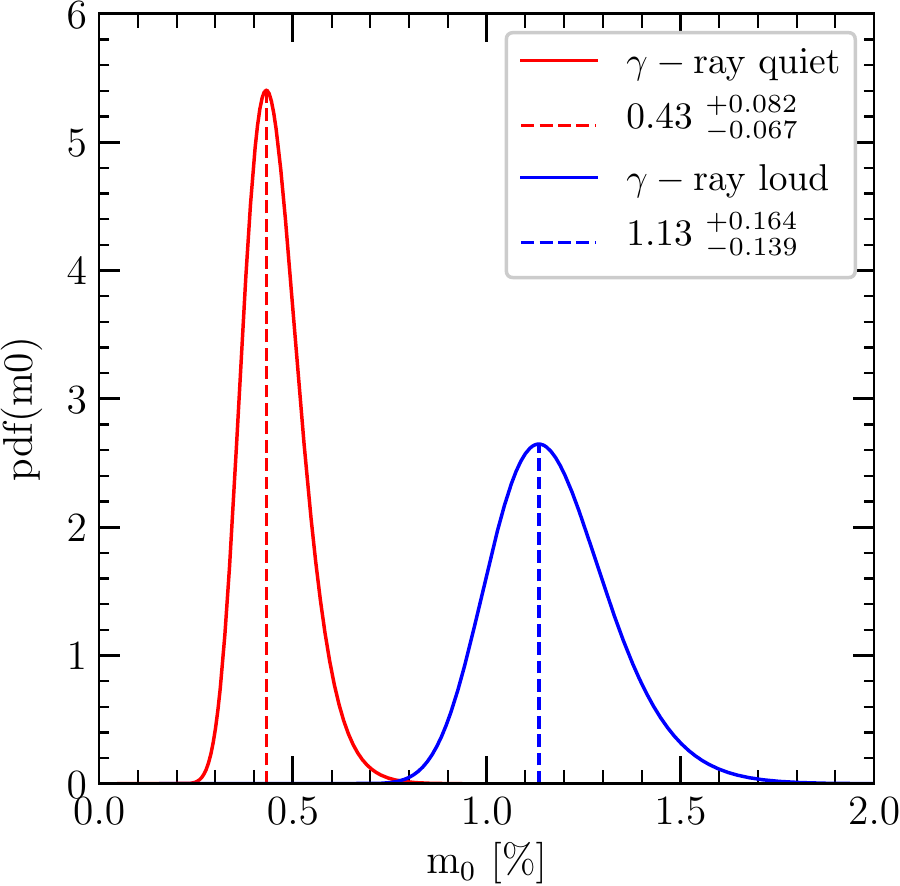}
     \includegraphics[width=0.46\textwidth]{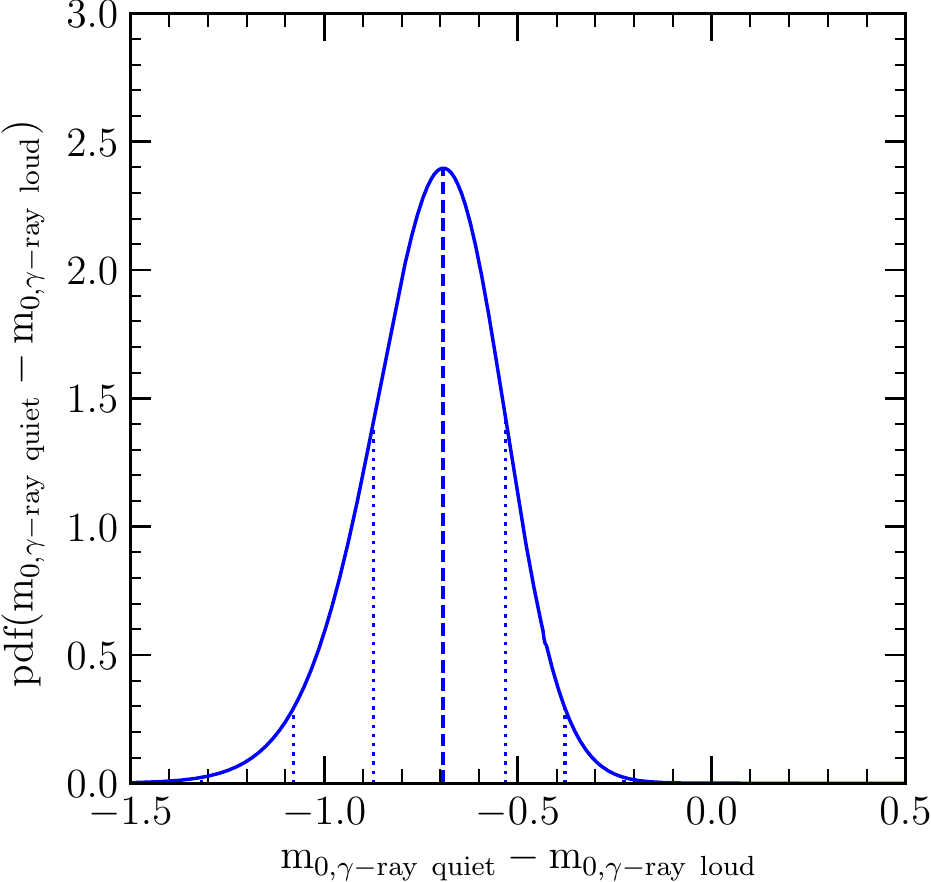}
     \caption{Similar to Figure \ref{fig:pop_flux} but for $\gamma$-ray quiet (red solid line) and $\gamma$-ray loud (blue solid line) sources. In the left panel the maximum-likelihood value and the associated 1$\sigma$ error are indicated in the legend. In the right panel the peak of the distribution ($-0.69^{+0.162}_{-0.179}$) is $\sim$4$\sigma$ away from zero.}
     \label{fig:pop_gamma}
\end{figure}

We thus divide our sample in two subsets, based on whether the source has been detected by Fermi LAT at a significance level high enough to warrant inclusion in the 3FGL catalog. As shown in Figure \ref{fig:pop_gamma}, these two subsamples reveal different properties: the $\gamma$-ray loud sources have, on average, an IDV amplitude almost a factor of four higher than $\gamma$-ray quiet ones. The~result is very significant statistically, with the maximum-likelihood difference being 4$\sigma$ away from 0, as indicated in the right panel of Figure \ref{fig:pop_gamma}.

We further investigated the possible relation between the integrated $\gamma$-ray photon flux and the radio flux density at 4.8\,GHz. The~$\gamma$-ray fluxes have been extracted from the 3FGL catalog \citep{Acero2015} and are averaged over the entire operational time of Fermi satellite. The~4.8\,GHz measurements are averaged over a few days and hence are most likely free of long -term variability. Our analysis showed that the two are likely correlated. In the case of photon fluxes in the range 100--300 MeV, the Spearman correlation coefficient turns out to be $r_s=0.46$ with $p=1.36\times10^{-4}$. In the case of fluxes in the range 1--100 GeV the correlation weakens, with $r_s$ being around 0.37 and a $p$-value of $3.15\times10^{-3}$. {These~numbers show that the radio flux density (which is an indicator of source compactness and brightness temperature) and $\gamma$-ray photon flux are significantly correlated.}

\subsection{Galactic Latitude} \label{sec:pop_b}

\begin{figure}[H]
\centering
     \includegraphics[width=0.45\textwidth]{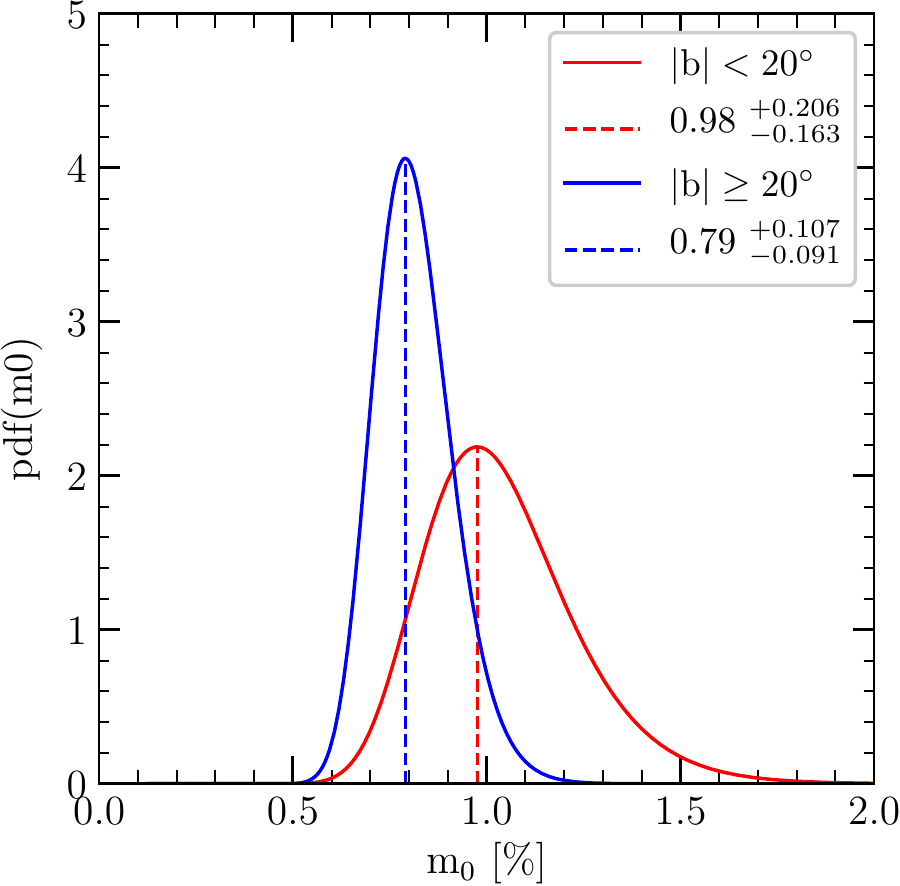}
     \includegraphics[width=0.46\textwidth]{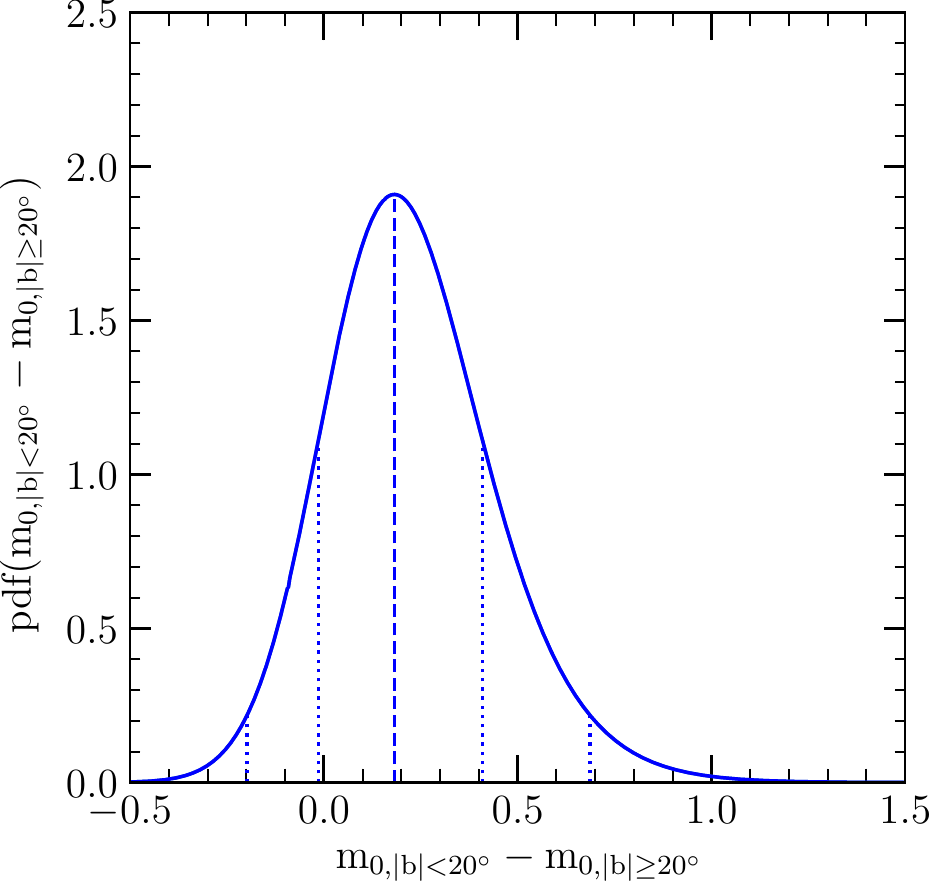}
     \caption{Similar to Figure \ref{fig:pop_flux} but for low (red solid line) and high (blue solid line) galactic latitude sources in our sample. In the left panel the maximum-likelihood value and the associated 1$\sigma$ error are indicated in the legend. In the right panel the peak of the distribution ($0.18^{+0.229}_{-0.194}$) is consistent with zero within $1\sigma$.}
     \label{fig:pop_b}
\end{figure}

\textls[-15]{We have also investigated the dependence of the IDV at the galactic latitude. In the case of ISS, a~galactic latitude dependence of IDV is anticipated, since the diffuse interstellar medium (ISM) is mostly distributed near the Galactic plane.
A contingency test by dividing the sample into low and high galactic latitude subsamples at $|b|=20^{\circ}$ reveals that although the low galactic latitude subsample on average exhibits marginally higher IDV amplitudes than that at high galactic latitude, the two distributions of $m_0$ are rather consistent with each other statistically. The~probability density for the difference between $m_0$ for the two subsamples is consistent with zero to within 1$\sigma$ (see Figure \ref{fig:pop_b}).}

In order to further verify this result, we then compare the IDV strength with the Galactic foreground emission measure (column density of the square of the electron density) as estimated from observations of H$\alpha$ emission (i.e., the Wisconsin H$\alpha$ Mapper Northern sky survey, WHAM \cite{Haffner2003}). The~integral H$\alpha$ intensity (in rayleighs) integrated over all velocities is believed to be proportional to the ISM emission measure in the line of sight, providing that the temperature of the emitting gas does not vary by a large percentage. By finding the integrated H$\alpha$ intensity from the WHAM nearest to each source, we are able to test the correlation between the IDV strength and the emission measure along the line of sight. We find no significant correlation between H$\alpha$ intensity and modulation index; the~Spearman correlation gives $r_s=0.180$ and $p=6.43\times10^{-2}$.

\textls[-15]{Previous studies with larger source samples suggest a significant variability dependence on Galactic latitude~(e.g.,~\cite[][]{Rickett2006, Lovell2008, Lazio2008}). A recent statistical study on AGN cores reveals that the effect of angular broadening by ISM scattering is significant only for sources at low galactic latitude (i.e.,~$|b|<10^{\circ}$)~\citep{Pushkarev2015}. Given the fact that only 12 of our sources are at low galactic latitudes  $|b|<10^{\circ}$, the lack of such a dependence in the current study might be simply ascribed to this small number. }

\section{Discussion} \label{sec:discuss}
\vspace{-6pt}
\subsection{Robustness of the Statistics}

A considerable portion of the sample was observed at multiple epochs. In that case, median values of source flux density and intrinsic modulation index are adopted for the statistical analysis presented in current study. To evaluate a possible bias introduced by the median-value selection approach for duplicate observations, both the minimum-value and maximum-value selection are tested. For~simplicity we hereafter refer to these two selection approaches as TMIN and TMAX, respectively.

\textls[-15]{The~distributions of intrinsic modulation index, for both TMIN and TMAX, show similar trends to that of the median-$\overline{m}$ selection, and can be characterized by an exponential distribution given in Equation~(\ref{eq:psd_im}), with $m_0$ = 0.62\% and 0.66\%, respectively. The~variability strengths for various subsamples are re-analyzed and compared by using the methodology demonstrated in Section~\ref{sec:pop_comp}. The~results are listed in Table~\ref{tab:pop_result_robust}. Though the estimated $m_0$ and $\Delta\,m_0$ values are systematically lower for TMIN and higher for TMAX, the significances are comparable. The~consistency between these results indicates that our previous findings on IDV dependence hold true even with `extreme' selection approaches on $\overline{m}$.}
\begin{table}[H]
\centering
\caption{Results of population comparisons with minimum- and maximum-$\overline{m}$ selection.}
\label{tab:pop_result_robust}
  \begin{tabular}{ccccccc}
  \toprule
   & \multicolumn{3}{c}{\textbf{TMIN}} & \multicolumn{3}{c}{\textbf{TMAX}} \\
\midrule
\textbf{Subsample}            & \boldmath {$m_0$}  \textbf{[\boldmath{\%}] }  & \boldmath {$\Delta\,m_0$ }   \textbf{[\boldmath{\%}] }       & \textbf{Significance}  & \boldmath {$m_0$}   \textbf{[\boldmath{\%}] } & \boldmath {$\Delta\,m_0$}      \textbf{[\boldmath{\%}] }     & \textbf{Significance}    \\
\midrule
$S_{4.8}<$\,1\,Jy     & $1.11_{-0.148}^{+0.178}$ & \multirow{2}{1.6cm}{$+0.65_{-0.172}^{+0.186}$}  & \multirow{2}{0.8cm}{$4\sigma$}
& $1.48_{-0.197}^{+0.237}$ & \multirow{2}{1.6cm}{$+0.99_{-0.236}^{+0.220}$}  & \multirow{2}{0.8cm}{$5\sigma$}    \\
$S_{4.8}\geq$\,1\,Jy  & $0.46_{-0.065}^{+0.078}$ &  &
& $0.48_{-0.066}^{+0.080}$ &  & \\
\midrule
$\alpha<$\,$-$0.1       & $0.58_{-0.082}^{+0.099}$ & \multirow{2}{1.6cm}{$-0.39_{-0.173}^{+0.160}$} & \multirow{2}{0.8cm}{$3\sigma$}
& $0.61_{-0.086}^{+0.104}$ & \multirow{2}{1.6cm}{$-0.68_{-0.222}^{+0.198}$} & \multirow{2}{0.8cm}{$4\sigma$} \\
$\alpha\geq$\,$-$0.1    & $0.97_{-0.126}^{+0.151}$ &  &
& $1.30_{-0.169}^{+0.202}$ &  &  \\
\midrule
$\gamma$-ray quiet    & $0.41_{-0.064}^{+0.079}$ & \multirow{2}{1.6cm}{$-0.62_{-0.168}^{+0.150}$} & \multirow{2}{0.8cm}{$4\sigma$}
& $0.45_{-0.069}^{+0.085}$ & \multirow{2}{1.6cm}{$-0.88_{-0.209}^{+0.184}$} & \multirow{2}{0.8cm}{$5\sigma$}  \\
$\gamma$-ray loud     & $1.05_{-0.128}^{+0.152}$ &  &
& $1.34_{-0.164}^{+0.194}$ &  &  \\
\midrule
$|b|<\,20^{\circ}$    & $0.89_{-0.148}^{+0.188}$ & \multirow{2}{1.6cm}{$+0.14_{-0.178}^{+0.212}$} & \multirow{2}{1cm}{<1$\sigma$}
& $1.17_{-0.195}^{+0.248}$ & \multirow{2}{1.6cm}{$+0.28_{-0.233}^{+0.267}$} & \multirow{2}{1cm}{$1\sigma$}   \\
$|b|\geq\,20^{\circ}$ & $0.74_{-0.086}^{+0.101}$ &  &
& $0.88_{-0.102}^{+0.119}$ &  &  \\
\bottomrule
\end{tabular}
\end{table}

\subsection{Variability Dependencies}

A strong dependence of IDV on source flux density is found in our sample. A similar effect was also observed in the MASIV survey \cite{Lovell2003}, where an increase in fractional amplitude of IDV with decreasing flux density was observed. The~result raised the possibility that the milliarcsecond-scale structures of the IDV sources may differ from those of non-IDVs, in the sense that the weaker sources are more compact. Consistent with this result, a direct comparison of milliarcsecond source structures between IDV/non-IDV sources showed that the former typically have smaller size than the latter \cite{Ojha2004d}. Moreover, the ISS-induced variability leads us to expect a flux density-dependent IDV, being common in compact objects but rare in objects with bright VLBI-scale jets~(e.g.,~\cite[][]{Quirrenbach1992, Lovell2003}). This is also the case for our sample, in which the fractional occurrence of IDV is $\sim$40$\%$ and $\sim$18$\%$ for the weak and strong subsamples, respectively. The~finding that IDV amplitude depends on the total flux density implies that the low flux-density sources identified in our sample are more compact.

It has long been suggested that flat-spectrum sources are more likely to show IDV \cite{Heeschen1987} than steep-spectrum sources. We narrow that conclusion by finding that in our selection of flat-spectrum sources, those with inverted spectra, on average, show stronger variability than sources with other spectral shapes. Since the flux density of inverted spectrum sources is more `core dominated', our~results strongly support the model that it is the most compact and core-dominant component of the blazars that causes IDV due to scintillation through the ISM.

We found a significant IDV dependence on $\gamma$-ray loudness in our sample. Prior to discussing the physical implications from this dependence, it is essential to discern the possible correlation between spectral indexes and $\gamma$-ray loudness, e.g., to test whether the $\gamma$-ray loud population is also dominated by inverted-spectrum sources.

The~K-S and A-D tests reveal that the distribution of the spectral index for $\gamma$-ray loud and $\gamma$-ray quiet sources is significantly different ($p_{K\text{-}S}=5.49\times10^{-4}$ and $p_{A\text{-}D}=1.34\times10^{-3}$ for K-S and A-D tests, respectively), with the $\gamma$-ray loud sample being dominated by the more inverted-spectrum sources. {As~mentioned above, the inverted-spectrum sources are dominated by optically thick compact components with higher brightness temperatures than their flat-spectrum counterparts.} For such sources the energy of radiative particles dominates over that of the magnetic field, and inverse-Compton (IC) scattering is therefore more efficient. This leads to an increased production of $\gamma$-rays {if IC
scattering is the dominant $\gamma$-ray emission process}. Another possibility is that the inverted spectrum (thus more core-dominated) sources are more strongly beamed~(e.g.,~\cite[][]{Hovatta2009}). In that case, as their apparent flux density is more strongly Doppler-boosted, for a given intrinsic brightness temperature they will have smaller angular sizes. Thus, they are more likely to scintillate and they are also more likely to have detectable $\gamma$-ray emission, which is dependent on beaming~(e.g.,~\cite[][]{Savolainen2010}). Furthermore, a positive correlation between the 4.8\,GHz radio flux density and the $\gamma$-ray photon flux is found, {which supports the view that it is the radio photons from the synchrotron branch of the spectral energy distribution (SED) that are up-scattered by the inverse-Compton process to the higher energies.}

{Our findings indicate a strong connection between the origin of radio and $\gamma$-ray emission. While~the radio emission of blazars is believed to be produced by synchrotron emission of relativistic electrons, which is Doppler-boosted, the origin of the $\gamma$-ray emission is still controversial, especially with regard to the target photon field and the location of the emission site~(see, e.g.,~\cite[][]{Maraschi1992, Dermer1992, Sikora1994}).
Besides~these leptonic models there are also models where the $\gamma$-rays originate from hadronic processes, i.e.,~ relativistic protons co-accelerated with the electrons~(e.g.,~\cite[][]{Mannheim1993}). In this case, rather strong magnetic fields would be required. Following \cite{Marscher1983}, the~typical magnetic field strength for the sources in our sample can be calculated. Due~to the lack of detailed spectral information for the scintillating component, we adopt a typical spectral turnover frequency near 15\,GHz from a recent statistical study of AGN jet compactness and brightness temperatures~(\cite[][]{Lee2016}, see Figure~1). For a source with an angular size of the order of the scattering size \mbox{$\theta$ = 0.05$\sim$0.25\,mas}, a~typical peak flux density \mbox{$\mathrm{S_m}$ = 1 Jy}, and Doppler factor $\delta$ = 10, one obtains a magnetic field strength of 1.7\,mG$\sim$1.1\,G, which would favor leptonic models, unless the turnover frequency or the Doppler factor would be much higher. Dedicated future studies will be necessary to further explore this~topic.}

The~presence of a relationship between ISS and galactic latitude has long been suggested from previous surveys~(e.g.,~\cite[][]{Rickett2006, Lovell2008, Lazio2008}). However, in this study only a marginal dependence was observed. Besides the smaller size of the sample as discussed in Section~\ref{sec:pop_b}, the lack of a clear correlation may indicate that {it is predominantly the intrinsic properties (e.g., angular size, core-dominance) of the blazars that determine how they scintillate, rather than properties of the distributed ISM. In other words, a sufficiently compact source is likely to show IDV, no matter through which part of the Galaxy it is observed. This statement excepts intra-hour variability~(e.g.,~\cite[][]{Kedziora-Chudczer1997, Dennett-Thorpe2000, Bignall2003}), an extremely rare ($\ll$1\%) phenomenon that requires an unusually nearby screen.}

\textls[-15]{{The~ISM is known to be highly inhomogeneous with small-scale discrete structures, such as photon-dominant regions (PDRs, e.g.,~\cite[][]{Hollenbach1997}), high-latitude clouds (HLCs, e.g.,~\cite[][]{Magnani1996}), local interstellar clouds (LICs, e.g.,~\cite[][]{Redfield2008}), and ionized flows driven by nearby hot stars~(e.g.,~\cite[][]{Walker2017}), etc. Such structures may dominate over the more diffuse ISM on the galactic latitude dependence of IDV: local structures within $\sim$100\,pc of the Sun can produce the strongest and most rapid IDV, due to the angular size of the first Fresnel zone scaling with the inverse square root of screen distance. Scintillation from more distant scattering screens is averaged out over the finite angular diameter of the source. The~high occurrence rate of IDV in compact radio sources indicates that for many AGN, the line-of-sight intersects such~inhomogeneities.}}

\subsection{Influence on SVLBI $T_B$ Measurements}

RadioAstron blazars that are strongly core-dominated should show significant IDV due to ISS, with the largest modulations expected near 5\,GHz (e.g.,~\cite[][]{Lovell2008}). To account for the effects of ISS during the RadioAstron observations, coordinated flux density monitoring over an extended period is necessary. To estimate the uncertainties of the $T_B$ measurements obtained from SVLBI and influenced by rapid flux variations, we assume that the source has a VLBI core component containing all the variable flux. We approximate this component with a circular Gaussian. The~relation between visibility and angular diameter of this Gaussian is expressed as \cite{Pearson1999}
\begin{equation}
 F(\rho)=\exp{\left(\dfrac{-(\pi \theta \rho)^2}{4\ln 2}\right)}
\end{equation}
where $\theta$ is the angular diameter and $\rho$ the baseline length. The~uncertainty of $\theta$ is given by
\begin{equation}
 \Delta\theta=-\frac{\overline{m}}{2}\cdot\frac{\sqrt{1+f_c^{-2}}}{\ln f_c}\cdot\theta
\end{equation}
in which $\overline{m}$ the intrinsic modulation index and $fc=\mathrm{S_{core}/S_{total}}$ the core-dominance. Analytically applying the error propagation, one obtains for the uncertainty of $T_B$
\begin{equation}
 \Delta T_B=\overline{m}\cdot\left(1+\dfrac{1+f_c^{-2}}{\ln^2f_c}\right)^{1/2}\cdot T_B
\end{equation}
which leads to an uncertainty of $\sim$$70\%$ in $T_B$ for a source which varies with $\overline{m}=10\%$ and whose core dominance is $f_c=0.8$.

\section{Summary and Conclusions} \label{sec:summary}

\textls[-15]{We presented statistical results based on five observing sessions of an IDV monitoring program with the Effelsberg 100-m radio telescope at 4.8\,GHz. The~overall statistics of the observed AGN showed that 31 out of 107 sources exhibited IDV, leading to an IDV detection rate of $\sim$30\%. The~IDV occurrence for $\gamma$-ray loud sources is $\sim$40\%, which is significantly higher than that for the $\gamma$-ray quiet ones.}

Moreover, with a maximum-likelihood approach we investigated the IDV dependence on various sources properties and on the galactic latitude. We found significant differences in the strength of IDV, dependent on the source flux density, spectral index and $\gamma$-ray loudness. The~results show that, weak (S\,<\,1\,Jy), inverted spectrum ($\alpha>-0.1$), or $\gamma$-ray loud sources, on average, exhibit significantly stronger IDV (significance $> 3 \sigma$). On the other hand, we did not find a significant dependence of IDV on the galactic latitude, which may suggest that it is predominantly the intrinsic properties (e.g., angular size, core-dominance) of the blazars that determine how they scintillate, rather than the directional dependence in the ISM. We estimate that for the blazars which show strong IDV, the~uncertainty in the observed VLBI brightness temperature can be as high as $\sim $$70\%$. A better physical understanding of these findings should become possible from direct size measurements through the ongoing RadioAstron space VLBI observations.

\vspace{6pt}

\acknowledgments{The~authors thank the anonymous referee for his comments, which helped to improve the paper. We also thank B. Boccardi for discussion and comments. This paper made use of data obtained with the 100-m telescope of the MPIfR (Max-Planck-Institut f\"ur Radioastronomie) at Effelsberg. This research was partially supported by the the Light in China's Western Region program (Grant No. 2015-XBQN-B-01, YBXM-2014-02, XBBS201324), the National Natural Science Foundation of China (NSFC, Grant No. 11503071, 11503072), the~National Basic Research Program of China (973 program, Grant No. 2015CB857100), Xinjiang Key Laboratory of Radio Astrophysics (Grant No. 2016D03020), and the China Scholarship Council (CSC, Grant No. 201704910392). Yuri Y. Kovalev acknowledges support by the government of the Russian Federation (agreement 05.Y09.21.0018) and the Alexander von Humboldt Foundation.}

\authorcontributions{J.L, H.B, T.P.K, X.L, A.K, Y.Y.K proposed the observations. Y.Y.K and K.V.S provided the RadioAstron schedule for this campaign. J.L, T.P.K and A.K performed the observations and conducted data calibration. H.B, T.P.K, X.L, E.A and J.A.Z contributed in the discussion and interpretation of the data. J.L wrote the paper.}

\conflictsofinterest{The~authors declare no conflict of interest.}

\appendixtitles{yes} 
\appendixsections{multiple} 
\appendix
\section{Results of IDV Monitoring with the Effelsberg 100\,m Telescope}

\footnotesize
\begin{center}
    \begin{longtable}{ccccccccccccccc}
\caption{Source properties and statistical results of Effelsberg monitoring (see Section \ref{sec:stat_overall} for detailed description).}
\label{tab:all_result}
  \endfirsthead
  \multicolumn{15}{c}{{\bfseries \tablename\ \thetable{} -- Cont.}} \\ \toprule
  Name    & Epoch & Flux density & $\overline{m}$     & $\chi_r^2$ & IDV & $b$ & $\alpha$ & $z$ & $\theta_5$ & $f_c$  & $\gamma$-ray \\
  &       &  [Jy]        &  [\%]              &            &     & [$^{\circ}$]  &      &           &   [mas]    &     & \\ \midrule
  \endhead  \midrule
  \endfoot \bottomrule
  \endlastfoot \toprule
  Name    & Epoch & Flux density & $\overline{m}$     & $\chi_r^2$ & IDV & $b$ & $\alpha$ & $z$ & $\theta_5$ & $f_c$  & $\gamma$-ray \\
  &       &  [Jy]        &  [\%]              &            &     & [$^{\circ}$]  &      &           &   [mas]    &     & \\
 \midrule
0010+405 & B & $0.976\pm0.002$  & $0.42_{-0.21}^{+0.24}$  &  1.586  &   & -21.44 & -0.24 & 0.255 & 0.70 & 0.78 &   \\
0014+813 & B & $1.642\pm0.002$  & $0.00_{-0.12}^{+0.12}$  &  0.133  &   &  18.80 & -0.01 & 3.366 & 0.63 & 0.75 &   \\
0016+731 & B & $2.106\pm0.005$  & $1.14_{-0.15}^{+0.23}$  &  4.485  & + &  10.73 & -0.13 & 1.781 & 0.57 & 0.79 &   \\
0059+581 & A & $3.517\pm0.011$  & $1.40_{-0.20}^{+0.31}$  &  2.607  & + &  -4.44 & -0.06 & 0.644 & 0.61 & 0.66 & + \\
0110+318 & C & $0.650\pm0.002$  & $0.00_{-0.61}^{+0.61}$  &  1.047  &   & -30.51 & -0.30 & 0.603 &      &      & + \\
         & D & $0.697\pm0.002$  & $1.32_{-0.19}^{+0.30}$  &  4.053  & + &        &       &       &      &      &   \\
         & E & $0.814\pm0.003$  & $0.00_{-0.42}^{+0.42}$  &  0.214  &   &        &       &       &      &      &   \\
0125+487 & B & $0.341\pm0.001$  & $0.61_{-0.18}^{+0.25}$  &  1.935  &   & -13.41 & +0.27 & 0.067 &      &      &   \\
0219+428 & D & $1.191\pm0.002$  & $0.38_{-0.24}^{+0.24}$  &  1.175  &   & -16.77 & -0.22 & 0.444 & 1.83 & 0.42 & + \\
0234+285 & E & $2.547\pm0.009$  & $0.00_{-0.46}^{+0.46}$  &  0.459  &   & -28.53 & -0.17 & 1.213 &      & 0.44 & + \\
0235+164 & E & $2.133\pm0.010$  & $1.19_{-0.49}^{+0.59}$  &  1.720  &   & -39.11 & +0.01 & 0.940 & 0.65 & 0.88 & + \\
0248+430 & D & $0.860\pm0.001$  & $0.53_{-0.19}^{+0.24}$  &  1.411  &   & -14.40 & -0.18 & 1.310 & 0.73 & 0.33 &   \\
0307+380 & B & $0.421\pm0.007$  & $7.08_{-0.99}^{+1.57}$  & 132.310 & + & -16.94 & +0.01 & 0.816 &      &      & + \\
         & C & $0.178\pm0.003$  & $5.85_{-0.91}^{+1.54}$  & 33.472  & + &        &       &       &      &      &   \\
         & D & $0.347\pm0.009$  & $13.02_{-1.63}^{+2.04}$ & 222.547 & + &        &       &       &      &      &   \\
         & E & $0.541\pm0.005$  & $3.56_{-0.57}^{+0.95}$  &  6.742  & + &        &       &       &      &      &   \\
0322+222 & D & $0.855\pm0.003$  & $1.51_{-0.22}^{+0.37}$  &  5.238  & + & -28.02 & -0.09 & 2.060 &      & 0.61 & + \\
0323+342 & C & $0.339\pm0.001$  & $1.05_{-0.36}^{+0.54}$  &  2.188  &   & -18.76 & -0.22 & 0.061 &      &      & + \\
         & D & $0.351\pm0.001$  & $0.57_{-0.23}^{+0.28}$  &  1.452  &   &        &       &       &      &      &   \\
0333+321 & C & $2.048\pm0.005$  & $0.00_{-0.34}^{+0.34}$  &  0.361  &   & -18.77 & -0.10 & 1.258 & 0.88 & 0.36 & + \\
         & D & $2.036\pm0.003$  & $0.41_{-0.20}^{+0.23}$  &  1.178  &   &        &       &       &      &      &   \\
         & E & $1.977\pm0.007$  & $0.00_{-0.71}^{+0.71}$  &  0.906  &   &        &       &       &      &      &   \\
0340+362 & D & $0.514\pm0.007$  & $6.42_{-0.78}^{+1.12}$  & 66.610  & + & -14.69 & -0.14 & 1.484 &      &      & + \\
0428+205 & B & $2.334\pm0.004$  & $0.00_{-0.18}^{+0.18}$  &  0.110  &   & -18.56 & -0.39 & 0.219 & 3.22 & 0.06 &   \\
0430+289 & E & $0.230\pm0.002$  & $3.69_{-0.58}^{+1.03}$  &  7.324  & + & -12.60 & -0.24 & 0.970 &      &      & + \\
0507+179 & B & $0.642\pm0.001$  & $0.17_{-0.17}^{+0.25}$  &  1.090  &   & -12.79 & +0.06 & 0.416 & 0.66 & 0.54 & + \\
0529+483 & B & $0.564\pm0.001$  & $0.00_{-0.26}^{+0.26}$  &  0.826  &   &   8.23 & -0.03 & 1.162 & 0.47 & 0.67 & + \\
         & E & $1.163\pm0.006$  & $1.84_{-0.34}^{+0.57}$  &  2.855  & + &        &       &       &      &      &   \\
0602+405 & B & $1.042\pm0.002$  & $0.00_{-0.36}^{+0.36}$  &  1.210  &   &   9.35 & -0.17 &       &      &      &   \\
0633+734 & B & $0.834\pm0.001$  & $0.00_{-0.14}^{+0.14}$  &  0.367  &   &  25.06 & -0.08 & 1.850 &      &      & + \\
0642+449 & B & $2.510\pm0.004$  & $0.00_{-0.17}^{+0.17}$  &  0.159  &   &  17.95 & -0.21 & 3.396 & 0.67 & 0.64 &   \\
0716+714 & A & $1.271\pm0.005$  & $2.00_{-0.25}^{+0.37}$  &  4.345  & + &  28.02 & +0.22 & 0.300 & 0.55 & 0.69 & + \\
         & B & $1.629\pm0.007$  & $2.21_{-0.24}^{+0.35}$  & 14.321  & + &        &       &       &      &      &   \\
         & C & $1.668\pm0.008$  & $2.79_{-0.28}^{+0.39}$  &  9.358  & + &        &       &       &      &      &   \\
         & D & $1.301\pm0.006$  & $2.97_{-0.29}^{+0.39}$  & 17.003  & + &        &       &       &      &      &   \\
         & E & $1.356\pm0.005$  & $1.92_{-0.29}^{+0.40}$  &  2.805  & + &        &       &       &      &      &   \\
0730+504 & D & $0.436\pm0.001$  & $0.54_{-0.22}^{+0.25}$  &  1.382  &   &  27.11 & -0.09 & 0.720 & 0.38 & 0.45 & + \\
0742+103 & E & $2.817\pm0.011$  & $0.00_{-0.50}^{+0.50}$  &  0.341  &   &  16.59 & -0.38 & 2.624 & 0.99 & 0.35 &   \\
0749+540 & B & $0.572\pm0.002$  & $1.47_{-0.20}^{+0.30}$  &  6.029  & + &  30.51 & -0.06 & 0.200 & 1.03 & 0.47 & + \\
         & C & $0.634\pm0.003$  & $2.15_{-0.27}^{+0.41}$  &  5.651  & + &        &       &       &      &      &   \\
         & D & $0.653\pm0.002$  & $1.46_{-0.17}^{+0.25}$  &  4.565  & + &        &       &       &      &      &   \\
0804+499 & A & $0.672\pm0.002$  & $0.00_{-0.34}^{+0.34}$  &  0.397  &   &  32.56 & +0.05 & 1.435 & 0.34 & 0.78 & + \\
         & B & $0.683\pm0.001$  & $0.00_{-0.24}^{+0.24}$  &  0.641  &   &        &       &       &      &      &   \\
0812+367 & C & $0.963\pm0.003$  & $0.17_{-0.17}^{+0.52}$  &  1.167  &   &  31.86 & -0.17 & 1.027 & 0.76 & 0.67 &   \\
0814+425 & A & $1.018\pm0.003$  & $0.60_{-0.30}^{+0.40}$  &  0.932  &   &  33.40 & -0.27 & 0.530 & 0.94 & 0.48 & + \\
         & B & $1.064\pm0.002$  & $0.44_{-0.22}^{+0.27}$  &  1.531  &   &        &       &       &      &      &   \\
         & E & $1.029\pm0.007$  & $5.07_{-0.42}^{+0.53}$  & 15.203  & + &        &       &       &      &      &   \\
0827+243 & E & $0.839\pm0.004$  & $0.00_{-0.57}^{+0.57}$  &  0.524  &   &  31.88 & +0.19 & 0.941 &      &      & + \\
0831+557 & A & $5.344\pm0.009$  & $0.00_{-0.28}^{+0.28}$  &  0.408  &   &  36.56 & -0.66 & 0.241 & 3.10 & 0.04 &   \\
         & B & $5.372\pm0.007$  & $0.00_{-0.14}^{+0.14}$  &  0.181  &   &        &       &       &      &      &   \\
         & C & $5.386\pm0.010$  & $0.00_{-0.20}^{+0.20}$  &  0.097  &   &        &       &       &      &      &   \\
         & D & $5.393\pm0.006$  & $0.00_{-0.12}^{+0.12}$  &  0.151  &   &        &       &       &      &      &   \\
0846+513 & C & $0.232\pm0.001$  & $0.00_{-0.37}^{+0.37}$  &  0.658  &   &  39.14 & +0.12 & 0.585 & 0.46 & 0.76 & + \\
0851+202 & D & $3.119\pm0.014$  & $1.77_{-0.26}^{+0.44}$  &  6.754  & + &  35.82 & +0.13 & 0.306 & 0.50 & 0.46 & + \\
         & E & $4.301\pm0.017$  & $0.00_{-0.57}^{+0.57}$  &  0.456  &   &        &       &       &      &      &   \\
0859+470 & A & $1.432\pm0.003$  & $0.00_{-0.26}^{+0.26}$  &  0.065  &   &  41.56 & -0.36 & 1.465 &      & 0.64 & + \\
         & B & $1.418\pm0.002$  & $0.00_{-0.17}^{+0.17}$  &  0.300  &   &        &       &       &      &      &   \\
0917+449 & A & $1.053\pm0.002$  & $0.00_{-0.34}^{+0.34}$  &  0.233  &   &  44.82 & -0.13 & 2.186 & 0.54 & 0.72 & + \\
0917+624 & A & $1.111\pm0.002$  & $0.00_{-0.20}^{+0.20}$  &  0.256  &   &  40.99 & -0.20 & 1.446 & 0.50 & 0.70 & + \\
0923+392 & A & $10.324\pm0.029$ & $0.00_{-0.36}^{+0.36}$  &  0.176  &   &  46.16 & -0.20 & 0.695 & 0.70 & 0.62 &   \\
0925+504 & A & $0.336\pm0.004$  & $3.87_{-0.60}^{+1.11}$  &  9.431  & + &  45.42 & -0.01 & 0.370 & 0.57 & 0.70 & + \\
         & B & $0.334\pm0.002$  & $2.99_{-0.38}^{+0.63}$  & 18.678  & + &        &       &       &      &      &   \\
         & C & $0.447\pm0.007$  & $7.17_{-0.92}^{+1.36}$  & 56.887  & + &        &       &       &      &      &   \\
         & D & $0.602\pm0.006$  & $5.79_{-0.61}^{+0.83}$  & 60.041  & + &        &       &       &      &      &   \\
         & E & $0.350\pm0.003$  & $4.54_{-0.55}^{+0.78}$  &  9.787  & + &        &       &       &      &      &   \\
0942+468 & C & $0.342\pm0.001$  & $0.00_{-0.30}^{+0.30}$  &  0.247  &   &  48.83 & -0.13 & 0.639 & 1.28 & 0.35 &   \\
0945+408 & A & $1.272\pm0.003$  & $0.00_{-0.32}^{+0.32}$  &  0.161  &   &  50.28 & -0.11 & 1.250 & 0.49 & 0.66 & + \\
0951+693 & C & $0.220\pm0.000$  & $0.00_{-0.36}^{+0.36}$  &  0.880  &   &  40.90 & -0.11 &       &      &      &   \\
0954+658 & A & $1.567\pm0.003$  & $0.00_{-0.25}^{+0.25}$  &  0.543  &   &  43.13 & -0.04 & 0.368 &      & 0.89 & + \\
         & E & $0.949\pm0.002$  & $0.00_{-0.32}^{+0.32}$  &  0.510  &   &        &       &       &      &      &   \\
0955+326 & A & $0.951\pm0.002$  & $0.00_{-0.30}^{+0.30}$  &  0.095  &   &  52.32 & -0.33 & 0.531 & 0.64 & 0.37 &   \\
0955+476 & A & $0.914\pm0.002$  & $0.26_{-0.26}^{+0.32}$  &  0.777  &   &  50.73 & -0.06 & 1.882 & 0.31 & 0.87 & + \\
1015+359 & A & $0.553\pm0.002$  & $0.64_{-0.30}^{+0.41}$  &  0.679  &   &  56.43 & +0.09 & 1.230 & 0.35 & 0.64 & + \\
1040+244 & E & $0.784\pm0.006$  & $2.39_{-0.44}^{+0.81}$  &  3.876  & + &  61.01 & +0.02 & 0.563 & 0.39 & 0.69 & + \\
1044+476 & C & $0.405\pm0.001$  & $0.00_{-0.33}^{+0.33}$  &  0.275  &   &  58.44 & -0.43 & 0.799 & 1.77 & 0.47 &   \\
1044+719 & A & $2.320\pm0.003$  & $0.00_{-0.24}^{+0.24}$  &  0.505  &   &  42.29 & -0.16 & 1.150 & 0.37 & 0.70 & + \\
         & E & $3.349\pm0.007$  & $0.00_{-0.34}^{+0.34}$  &  0.574  &   &        &       &       &      &      &   \\
1053+815 & A & $0.381\pm0.001$  & $0.99_{-0.19}^{+0.26}$  &  1.122  &   &  34.75 & -0.01 & 0.706 &      &      & + \\
1101+384 & A & $0.634\pm0.002$  & $0.00_{-0.40}^{+0.40}$  &  0.634  &   &  65.03 & -0.10 & 0.030 &      & 0.85 & + \\
         & E & $0.617\pm0.002$  & $0.00_{-0.56}^{+0.56}$  &  0.569  &   &        &       &       &      &      &   \\
1123+264 & A & $1.248\pm0.003$  & $0.00_{-0.30}^{+0.30}$  &  0.212  &   &  70.89 & -0.03 & 2.352 & 0.34 & 0.71 &   \\
         & B & $1.250\pm0.002$  & $0.00_{-0.33}^{+0.33}$  &  0.829  &   &        &       &       &      &      &   \\
         & C & $1.227\pm0.003$  & $0.00_{-0.37}^{+0.37}$  &  0.390  &   &        &       &       &      &      &   \\
         & D & $1.253\pm0.003$  & $0.56_{-0.22}^{+0.30}$  &  1.491  &   &        &       &       &      &      &   \\
         & E & $1.225\pm0.005$  & $0.00_{-0.46}^{+0.46}$  &  0.185  &   &        &       &       &      &      &   \\
1125+596 & A & $0.692\pm0.004$  & $3.28_{-0.37}^{+0.55}$  &  9.795  & + &  54.67 & -0.02 & 1.795 & 0.24 & 0.78 & + \\
         & C & $0.773\pm0.006$  & $4.29_{-0.44}^{+0.61}$  & 20.143  & + &        &       &       &      &      &   \\
         & D & $0.807\pm0.008$  & $6.28_{-0.60}^{+0.76}$  & 76.008  & + &        &       &       &      &      &   \\
         & E & $0.840\pm0.005$  & $3.66_{-0.41}^{+0.58}$  &  8.054  & + &        &       &       &      &      &   \\
1144+402 & A & $1.797\pm0.004$  & $0.00_{-0.45}^{+0.45}$  &  0.489  &   &  71.47 & -0.00 & 1.090 & 0.23 & 0.73 & + \\
         & D & $1.327\pm0.002$  & $0.00_{-0.28}^{+0.28}$  &  0.703  &   &        &       &       &      &      &   \\
1150+497 & D & $1.377\pm0.002$  & $0.00_{-0.20}^{+0.20}$  &  0.462  &   &  64.98 & -0.10 & 0.334 & 0.73 & 0.50 & + \\
1150+812 & B & $1.172\pm0.001$  & $0.00_{-0.12}^{+0.12}$  &  0.178  &   &  35.84 & -0.05 & 1.250 &      & 0.51 &   \\
1156+295 & A & $1.167\pm0.003$  & $0.00_{-0.29}^{+0.29}$  &  0.140  &   &  78.37 & -0.17 & 0.725 & 0.61 & 0.77 & + \\
         & B & $1.153\pm0.004$  & $1.24_{-0.22}^{+0.37}$  &  5.142  & + &        &       &       &      &      &   \\
         & C & $1.442\pm0.005$  & $0.66_{-0.38}^{+0.50}$  &  1.548  &   &        &       &       &      &      &   \\
         & D & $1.521\pm0.003$  & $0.72_{-0.18}^{+0.28}$  &  1.906  &   &        &       &       &      &      &   \\
         & E & $1.645\pm0.005$  & $0.00_{-0.71}^{+0.71}$  &  0.967  &   &        &       &       &      &      &   \\
1214+588 & B & $0.732\pm0.001$  & $0.00_{-0.14}^{+0.14}$  &  0.348  &   &  57.97 & -0.13 & 2.551 & 0.89 & 0.36 &   \\
         & C & $0.763\pm0.001$  & $0.00_{-0.21}^{+0.21}$  &  0.289  &   &        &       &       &      &      &   \\
1219+285 & D & $0.722\pm0.001$  & $0.00_{-0.31}^{+0.31}$  &  0.695  &   &  83.29 & -0.09 & 0.102 & 0.58 & 0.36 & + \\
1222+216 & E & $2.541\pm0.010$  & $0.00_{-0.46}^{+0.46}$  &  0.056  &   &  81.66 & -0.27 & 0.432 &      & 0.78 & + \\
1333+589 & B & $0.701\pm0.002$  & $1.05_{-0.15}^{+0.22}$  &  3.800  & + &  57.48 & +0.16 & 0.570 & 1.44 & 0.20 &   \\
         & C & $0.704\pm0.001$  & $0.70_{-0.24}^{+0.30}$  &  1.528  &   &        &       &       &      &      &   \\
1357+769 & A & $0.392\pm0.001$  & $1.32_{-0.22}^{+0.34}$  &  1.741  &   &  39.77 & -0.10 & 1.585 & 0.42 & 0.82 & + \\
         & E & $0.196\pm0.001$  & $4.28_{-0.39}^{+0.48}$  &  8.218  & + &        &       &       &      &      &   \\
1404+286 & D & $2.154\pm0.004$  & $0.00_{-0.19}^{+0.19}$  &  0.200  &   &  73.25 & +0.00 &       & 0.94 & 0.40 &   \\
1417+385 & C & $0.618\pm0.003$  & $1.19_{-0.30}^{+0.52}$  &  2.528  & + &  68.38 & +0.02 & 1.830 & 0.48 & 0.86 & + \\
         & D & $0.543\pm0.002$  & $1.02_{-0.22}^{+0.32}$  &  2.597  & + &        &       &       &      &      &   \\
1435+638 & A & $1.493\pm0.002$  & $0.00_{-0.19}^{+0.19}$  &  0.106  &   &  49.73 & -0.36 & 2.068 & 1.11 & 0.33 &   \\
1520+319 & B & $0.471\pm0.001$  & $0.00_{-0.41}^{+0.41}$  &  1.121  &   &  57.02 & -0.01 & 1.487 & 0.32 & 0.84 & + \\
1547+507 & C & $0.782\pm0.006$  & $3.93_{-0.46}^{+0.66}$  & 15.284  & + &  49.06 & -0.02 & 2.171 & 1.03 & 0.56 &   \\
         & D & $0.792\pm0.005$  & $3.29_{-0.38}^{+0.57}$  & 18.832  & + &        &       &       &      &      &   \\
         & E & $0.806\pm0.007$  & $4.06_{-0.56}^{+0.87}$  &  7.844  & + &        &       &       &      &      &   \\
1617+229 & D & $0.652\pm0.003$  & $1.75_{-0.28}^{+0.49}$  &  6.534  & + &  43.03 & -0.05 & 1.987 & 0.29 & 0.84 &   \\
1633+382 & A & $2.683\pm0.006$  & $0.00_{-0.30}^{+0.30}$  &  0.280  &   &  42.34 & -0.14 & 1.813 & 1.14 & 0.18 & + \\
         & B & $2.727\pm0.004$  & $0.00_{-0.19}^{+0.19}$  &  0.268  &   &        &       &       &      &      &   \\
         & C & $2.365\pm0.006$  & $0.00_{-0.42}^{+0.42}$  &  0.696  &   &        &       &       &      &      &   \\
         & E & $2.207\pm0.005$  & $0.00_{-0.24}^{+0.24}$  &  0.148  &   &        &       &       &      &      &   \\
1638+398 & C & $1.328\pm0.003$  & $0.00_{-0.31}^{+0.31}$  &  0.270  &   &  41.42 & -0.08 & 1.660 & 0.28 & 0.82 & + \\
1641+399 & A & $5.975\pm0.012$  & $0.00_{-0.24}^{+0.24}$  &  0.157  &   &  40.95 & -0.20 & 0.593 & 1.28 & 0.24 & + \\
1642+264 & D & $0.084\pm0.000$  & $0.70_{-0.66}^{+0.57}$  &  1.286  &   &  38.40 & -0.21 &       &      &      &   \\
1642+690 & E & $1.860\pm0.005$  & $0.00_{-0.30}^{+0.30}$  &  0.171  &   &  36.62 & -0.12 & 0.751 & 0.75 & 0.50 &   \\
1714+219 & E & $0.534\pm0.003$  & $0.00_{-0.60}^{+0.60}$  &  0.139  &   &  30.22 & -0.17 & 0.358 &      &      &   \\
1726+455 & C & $1.217\pm0.004$  & $0.44_{-0.34}^{+0.37}$  &  1.275  &   &  33.28 & +0.00 & 0.717 & 0.34 & 0.87 & + \\
         & D & $1.559\pm0.003$  & $0.00_{-0.29}^{+0.29}$  &  0.646  &   &        &       &       &      &      &   \\
1751+288 & B & $1.684\pm0.002$  & $0.00_{-0.16}^{+0.16}$  &  0.123  &   &  24.46 & +0.08 & 1.115 & 0.82 & 0.86 &   \\
         & C & $1.731\pm0.005$  & $0.00_{-0.43}^{+0.43}$  &  0.627  &   &        &       &       &      &      &   \\
1758+388 & B & $0.738\pm0.001$  & $0.00_{-0.26}^{+0.26}$  &  0.817  &   &  26.03 & -0.16 & 2.092 & 0.30 & 0.87 &   \\
         & C & $0.697\pm0.002$  & $0.00_{-0.39}^{+0.39}$  &  0.550  &   &        &       &       &      &      &   \\
1807+279 & B & $0.653\pm0.001$  & $0.00_{-0.17}^{+0.17}$  &  0.198  &   &  20.97 & -0.39 & 1.760 &      &      &   \\
         & D & $0.624\pm0.002$  & $0.72_{-0.19}^{+0.32}$  &  1.843  &   &        &       &       &      &      &   \\
1807+698 & A & $1.799\pm0.003$  & $0.00_{-0.27}^{+0.27}$  &  0.425  &   &  29.17 & -0.08 & 0.051 & 0.83 & 0.64 & + \\
         & B & $1.803\pm0.002$  & $0.00_{-0.14}^{+0.14}$  &  0.493  &   &        &       &       &      &      &   \\
1817+387 & C & $0.159\pm0.001$  & $1.69_{-0.32}^{+0.56}$  &  3.412  & + &  22.46 & +0.56 & 0.540 &      &      &   \\
         & E & $0.219\pm0.001$  & $0.00_{-0.62}^{+0.62}$  &  0.530  &   &        &       &       &      &      &   \\
1823+568 & B & $1.128\pm0.002$  & $0.99_{-0.14}^{+0.22}$  &  3.527  & + &  26.08 & -0.03 & 0.664 & 0.43 & 0.82 & + \\
1842+681 & C & $0.662\pm0.001$  & $0.54_{-0.27}^{+0.27}$  &  1.322  &   &  25.85 & -0.09 & 0.472 &      & 0.77 & + \\
1843+356 & C & $0.784\pm0.002$  & $0.00_{-0.34}^{+0.34}$  &  0.348  &   &  16.53 & -0.22 & 0.764 &      &      &   \\
1846+322 & C & $0.572\pm0.003$  & $1.29_{-0.31}^{+0.52}$  &  2.781  & + &  14.71 & -0.12 & 0.798 & 4.49 &      & + \\
         & D & $0.506\pm0.002$  & $1.06_{-0.21}^{+0.33}$  &  2.796  & + &        &       &       &      &      &   \\
1849+670 & A & $1.226\pm0.003$  & $0.95_{-0.20}^{+0.27}$  &  1.209  &   &  25.04 & +0.19 & 0.657 & 0.54 & 0.86 & + \\
         & B & $1.257\pm0.002$  & $0.84_{-0.13}^{+0.20}$  &  2.991  & + &        &       &       &      &      &   \\
1850+402 & E & $0.713\pm0.003$  & $0.00_{-0.42}^{+0.42}$  &  0.113  &   &  16.97 & -0.07 & 2.120 &      &      &   \\
1923+210 & C & $1.406\pm0.004$  & $0.00_{-0.33}^{+0.33}$  &  0.256  &   &   2.26 & -0.12 &       & 1.29 & 0.21 &   \\
1926+611 & A & $0.960\pm0.007$  & $3.57_{-0.42}^{+0.64}$  & 11.904  & + &  19.45 & -0.04 &       & 0.71 & 0.64 & + \\
         & E & $0.916\pm0.002$  & $0.00_{-0.43}^{+0.43}$  &  0.722  &   &        &       &       &      &      &   \\
1928+738 & B & $3.032\pm0.003$  & $0.00_{-0.16}^{+0.16}$  &  0.567  &   &  23.54 & -0.24 & 0.302 & 1.01 & 0.50 &   \\
1954+513 & C & $1.469\pm0.003$  & $0.00_{-0.33}^{+0.33}$  &  0.771  &   &  11.76 & -0.22 & 1.220 & 0.66 & 0.67 &   \\
         & E & $1.370\pm0.004$  & $0.00_{-0.55}^{+0.55}$  &  0.771  &   &        &       &       &      &      &   \\
2005+403 & C & $2.331\pm0.006$  & $0.00_{-0.34}^{+0.34}$  &  0.389  &   &   4.30 & -0.12 & 1.736 &      &      &   \\
2005+642 & A & $0.487\pm0.001$  & $0.00_{-0.41}^{+0.41}$  &  0.553  &   &  16.73 & +0.38 & 1.574 &      &      &   \\
2007+777 & B & $0.700\pm0.002$  & $1.27_{-0.16}^{+0.24}$  &  5.390  & + &  22.73 & -0.19 & 0.342 & 0.72 & 0.64 & + \\
         & D & $0.761\pm0.003$  & $2.35_{-0.23}^{+0.32}$  & 10.748  & + &        &       &       &      &      &   \\
2010+723 & A & $0.878\pm0.002$  & $1.10_{-0.21}^{+0.30}$  &  1.923  &   &  20.18 & -0.27 &       &      &      &   \\
2013+370 & E & $3.890\pm0.014$  & $0.00_{-0.43}^{+0.43}$  &  0.158  &   &   1.22 & -0.03 &       &      &      & + \\
2021+614 & D & $3.107\pm0.003$  & $0.00_{-0.16}^{+0.16}$  &  0.559  &   &  13.78 & -0.19 & 0.227 & 0.79 & 0.36 &   \\
2022+542 & C & $0.664\pm0.001$  & $0.00_{-0.26}^{+0.26}$  &  0.570  &   &   9.66 & -0.90 &       &      & 0.30 &   \\
2023+760 & A & $0.714\pm0.003$  & $1.43_{-0.22}^{+0.35}$  &  2.677  & + &  21.13 & +0.07 & 0.594 &      &      & + \\
2107+353 & C & $1.093\pm0.003$  & $0.00_{-0.51}^{+0.51}$  &  0.852  &   &  -8.35 & +0.13 & 0.202 & 1.48 & 0.18 & + \\
2136+141 & E & $2.438\pm0.011$  & $0.00_{-0.52}^{+0.52}$  &  0.293  &   & -27.54 & -0.14 & 2.427 & 0.60 & 0.66 &   \\
2155+312 & B & $0.418\pm0.001$  & $1.27_{-0.20}^{+0.33}$  &  4.932  & + & -18.24 & +0.02 & 1.486 &      &      & + \\
         & E & $0.645\pm0.004$  & $1.69_{-0.42}^{+0.79}$  &  2.416  &   &        &       &       &      &      &   \\
2200+420 & D & $3.060\pm0.005$  & $0.56_{-0.17}^{+0.24}$  &  1.473  &   & -10.44 & -0.08 & 0.069 & 0.37 & 0.45 & + \\
         & E & $2.146\pm0.007$  & $0.00_{-0.62}^{+0.62}$  &  0.793  &   &        &       &       &      &      &   \\
2201+315 & C & $2.465\pm0.006$  & $0.00_{-0.28}^{+0.28}$  &  0.148  &   & -18.78 & -0.09 & 0.295 & 0.94 & 0.45 & + \\
2209+236 & C & $0.796\pm0.003$  & $0.45_{-0.37}^{+0.39}$  &  1.310  &   & -26.09 & +0.13 & 1.125 &      & 0.85 & + \\
         & D & $1.059\pm0.003$  & $1.07_{-0.20}^{+0.34}$  &  3.085  & + &        &       &       &      &      &   \\
2223+210 & C & $1.547\pm0.004$  & $0.00_{-0.35}^{+0.35}$  &  0.444  &   & -30.09 & -0.24 & 1.959 & 1.26 & 0.17 &   \\
         & D & $1.593\pm0.003$  & $0.00_{-0.18}^{+0.18}$  &  0.143  &   &        &       &       &      &      &   \\
2230+114 & E & $4.235\pm0.017$  & $0.00_{-0.46}^{+0.46}$  &  0.117  &   & -38.58 & -0.20 & 1.037 & 1.04 & 0.40 & + \\
2309+454 & B & $0.699\pm0.003$  & $1.69_{-0.23}^{+0.37}$  &  7.917  & + & -13.70 & +0.13 & 1.447 &      &      &   \\
         & D & $0.530\pm0.003$  & $2.40_{-0.30}^{+0.47}$  & 10.230  & + &        &       &       &      &      &   \\
         & E & $0.550\pm0.003$  & $1.28_{-0.39}^{+0.61}$  &  1.766  &   &        &       &       &      &      &   \\
2351+456 & B & $1.142\pm0.002$  & $0.00_{-0.15}^{+0.15}$  &  0.152  &   & -15.85 & -0.26 & 1.992 &      & 0.50 & + \\
\end{longtable}
\end{center}
\normalsize

\reftitle{References}


\end{document}